\documentclass[11pt,a4paper]{article}

\usepackage{SupportingFiles/jheppubBlind}
\usepackage[utf8]{inputenc}
\usepackage[british]{babel}
\usepackage{multirow}
\usepackage[dvipsnames]{xcolor}
\usepackage{arydshln}
\usepackage{overpic}
\usepackage{etoolbox}
\usepackage{mathtools}
\usepackage{latexsym,exscale,amssymb,amsmath,amsthm}
\usepackage{sidecap}
\usepackage{braket}
\usepackage{dsfont}
\usepackage{braket}
\usepackage{tikz}
\usepackage{hyperref}
\usepackage{chngcntr}

\hypersetup{
    colorlinks,
    citecolor=red,
    filecolor=black,
    linkcolor=blue,
    urlcolor=blue
}
\counterwithin*{section}{part}
\numberwithin{equation}{section}
\numberwithin{table}{section}
\numberwithin{figure}{section}
\appto\appendix{\addtocontents{toc}{\protect\setcounter{tocdepth}{1}}}

\definecolor{highlightColour}{RGB}{0,0,150}
\definecolor{lightGrey}{RGB}{225,225,225}
\newcommand{\vect}[1]{\boldsymbol{#1}}

\newcommand{\R}{\mathbb{R}}

\newcommand{\Z}{\mathbb{Z}}
\newcommand{\N}{\mathcal{N}}
\newcommand{\A}{\mathcal{A}}
\newcommand{\C}{\mathbb{C}}
\newcommand{\X}{\mathcal{X}}
\newcommand{\unit}{\mathbf{1}}
\newcommand{\plAlt}[1]{p_{#1}}
\newcommand{\plTrop}[1]{w_{#1}}
\newcommand{\pl}[1]{\left<#1\right>}
\newcommand{\val}[1]{\text{val}\left(#1\right)}
\newcommand{\T}{\mathbb{T}}
\newcommand{\E}{E}
\newcommand{\g}{\vect{g}}
\newcommand{\hsline}[1]{\frac{\,\,\,\,#1\,\,\,\,}{}}
\newcommand{\trop}{\text{Tr}}
\newcommand{\G}[1]{\operatorname{Gr}({#1})}
\newcommand{\TG}[1]{\operatorname{Tr}({#1})}
\newcommand{\TPTG}[1]{\operatorname{Tr}_+({#1})}
\newcommand{\Conf}[1]{\widetilde{\operatorname{Gr}}({#1})}
\newcommand{\PConf}[1]{\widetilde{\operatorname{Gr}}_+({#1})}
\newcommand{\TC}[1]{\widetilde{\operatorname{Tr}}({#1})}
\newcommand{\TPTC}[1]{\widetilde{\operatorname{Tr}}_+({#1})}
\newtheorem{theorem}{Theorem}[section]

\theoremstyle{definition}

\newtheorem{example}[theorem]{Example}

\newcommand{\be}{\begin{equation}}
\newcommand{\ee}{\end{equation}}

\title{How tropical are seven- and eight-particle amplitudes?}
\author{Niklas Henke, Georgios Papathanasiou}
\affiliation{DESY Theory Group, DESY Hamburg, Notkestraße 85, D-22607 Hamburg}
\emailAdd{niklas.henke@desy.de}
\emailAdd{georgios.papathanasiou@desy.de}
\preprint{DESY 19-229}

\abstract{We study tropical Grassmannians Tr$(k,n)$ in relation to cluster algebras, and assess their applicability to $n$-particle amplitudes for $n=7,8$. In $\mathcal{N}=4$ super Yang-Mills theory, we first show that while the totally positive part of Tr$(4,7)$ may encompass the iterated discontinuity structure of the seven-point Maximally Helicity Violating (MHV) amplitude, it is too small for the Next-to-MHV helicity configuration. Then, using Tr$(4,8)$ we propose a finite set of 356 cluster $\mathcal{A}$-coordinates expected to contain the rational symbol letters of the eight-particle MHV amplitude, and discuss  how the remaining square-root letters may be obtained from limits of infinite mutation sequences. Finally, we use a triangulation of the totally positive part of Tr$(3,8)$ to obtain the associated generalised biadjoint scalar amplitude in a form containing a near-minimal amount of spurious poles.
}

\begin{document}

\maketitle

\section{Introduction}
Scattering amplitudes in perturbative quantum field theory have since long been the source of ever-renewing interplay between beautiful mathematics and realistic applications. This is arguably even more so the case for the simplest interacting four-dimensional gauge-theory, maximally supersymmetric or $\N=4$ super Yang-Mills (SYM) in the planar limit. For example, while the algebraic structure of a relevant class of functions known as multiple polylogarithms (MPLs) was well-established in the mathematics literature for some time \cite{Gonch3,Gonch2,Brown:2011ik}, it was its application for the first time in a physics context in this theory \cite{Goncharov:2010jf}, which led to an explosion of important phenomenological two-loop results, starting with \cite{Duhr:2012fh}.

While more complicated functions are certainly known to appear in generic quantum field theories at two loops~\cite{SABRY1962401} and beyond, arguments at the level of the integrand \cite{ArkaniHamed:2012nw} suggest that (appropriately normalised \cite{Bern:2005iz, Alday:2009dv, Yang:2010as,Dixon:2014iba}) amplitudes in the maximally helicity violating (MHV) or next-to-MHV (NMHV) configuration in $\N=4$ SYM are MPLs of weight $2L$ at $L$ loops for any number of external legs. A function $F$ is defined to be an MPL
of weight $n$ if its total differential obeys
\be 
dF = \sum_{\phi_\beta} F^{\phi_\beta} d\ln \phi_\beta\,,
\label{dFPhi}
\ee
such that $F^{\phi_\alpha}$ is a MPL of weight $n-1$ and so on, with the above recursive definition terminating with the usual logarithms
on the left-hand side at weight one, and (weight zero) rational numbers accompanying the 
total differentials on the right-hand side. The \emph{symbol} \cite{Goncharov:2010jf} is a convenient tool encapsulating this recursive definition by mapping $F$ to an $n$-fold tensor product, where the set $\phi_\beta$ in \eqref{dFPhi} is placed on the rightmost factor, the $d\log$ argument of the analogous relation for $dF^{\phi_\alpha}$ is placed on the next-to-rightmost factor, and so forth. The union of all $d\log$ arguments such as $\phi_\beta$ from all the tensor product factors is the \emph{symbol alphabet}, and it evidently encodes the singularity and discontinuity structure of the function $F$.

At low orders, the symbol alphabet (in fact, the entire amplitude) may be obtained by direct Feynman diagram \cite{DelDuca:2009au,DelDuca:2010zg} or symmetry-related \cite{CaronHuot:2011ky,Golden:2014xqf} computations. More generally, however, there is evidence \cite{Golden2014} that the symbol aphabet is dictated by another intriguing mathematical object known as a (Grassmannian) cluster algebra \cite{1021.16017,1054.17024,CAIII,CAIV}. In particular, this object is naturally defined in the space of kinematics of the $n$-particle amplitude, which coincides with the Gr$(4,n)$ \emph{Grassmannian}, thanks to the dual conformal symmetry~\cite{Drummond:2007au,Drummond:2006rz,Bern:2006ew,Bern:2007ct,Alday:2007he} of the theory\footnote{In particular, this dual conformal symmetry also implies that only normalised amplitudes with $n\ge 6$ have nontrivial kinematic dependence, and are therefore of interest.}. More precisely, the space of kinematics is a certain quotient of the Grassmannian, the \emph{configuration space} (of $n$ points in complex projective space $\mathbb{P}^3$) $\Conf{4,n}$, as can be most easily seen by using kinematic variables known as momentum twistors~\cite{Hodges:2009hk}, which conveniently realise the aforementioned symmetry (a pedagogical introduction on momentum twistors is contained in \cite{ArkaniHamed:2010gh}).

More recently, it has been realised that cluster algebras also provide more `local' information. Namely they provide not only the entire symbol alphabet, but also which \emph{letters} thereof are allowed to appear in adjacent entries of the symbol \cite{Drummond:2017ssj}. This property of \emph{cluster adjacency} has also been extended to the rational factors that may be present in the amplitudes \cite{Drummond:2018dfd}, for more recent applications see \cite{Drummond:2018caf,Golden:2019kks,Mago:2019waa,Lukowski:2019sxw}. Very interestingly, while the direct physical origin of cluster adjacency remains obscure, for the space of functions with physical branch cuts containing six- and seven-particle amplitudes, cluster adjacency is equivalent to the physically more transparent extended Steinmann relations\footnote{Cluster adjacency/extended Steinmann relations do not hold in the original, BDS normalisation \cite{Bern:2005iz} of the amplitude. However they are obeyed in the more recent, BDS-like normalisation \cite{Alday:2009dv, Yang:2010as,Dixon:2014iba} (as also reviewed in \cite{Dixon:2016nkn}), and it is the latter that we will use in section \ref{sec:tptg47}. } \cite{YorgosAmps17,Caron-Huot:2018dsv,Caron-Huot:2019bsq}. 

Thanks to the knowledge of the symbol alphabet, cluster adjacency/extended Steinmann relations or more generally the analytic structure of the amplitude, a \emph{bootstrap method} has been developed, which enables its construction without ever having to resort to Feynman diagrams. It at most requires information on the behaviour of the amplitude in certain kinematic regions, such as the multi-Regge~\cite{Bartels:2008sc,Fadin:2011we,Bartels:2011ge,Lipatov:2012gk,Dixon:2012yy,Bartels:2013jna,Basso:2014pla,Drummond:2015jea,DelDuca:2016lad,DelDuca:2018hrv,Marzucca:2018ydt,DelDuca:2019tur} or collinear limit~\cite{Alday:2010ku,Basso:2013vsa,Basso:2013aha,Papathanasiou:2013uoa,Basso:2014koa,Papathanasiou:2014yva,Basso:2014nra,Belitsky:2014sla,Belitsky:2014lta}, which may be independently and relatively simply be obtained by other means, and has been successfully applied through seven loops for the six-particle amplitude~\cite{Dixon:2011pw,Dixon:2011nj,Dixon:2013eka,Dixon:2014voa,Dixon:2014xca,Dixon:2014iba,Dixon:2015iva,Caron-Huot:2016owq,Caron-Huot:2019vjl}
and through four loops for the (symbol of the) seven-particle amplitude~\cite{Drummond:2014ffa,Dixon:2016nkn,Drummond:2018caf}. In principle, the problem of determining the amplitude (symbol) with $n=6,7$ legs in $\N=4$ SYM is thus solved, subject to limitations in (still orders of magnitude less compared to Feynman diagrams) computational power.

Despite this progress, a serious conceptual and practical obstacle prevents its straightforward generalisation to amplitudes with $n\ge 8$: In this case, the corresponding Gr$(4,n)$ Grassmannian cluster algebra becomes infinite, even though the amplitude can only have a finite number of symbol letters at fixed loop order. While the symbol alphabet (of size possibly increasing with the loop order) may still be contained in the cluster algebra, the fact that we have infinite possibilities to choose from means that in reality we know very little about the analytic structure of the amplitude, and certainly not enough to bootstrap it.

Hope for overcoming this obstacle has recently been raised thanks to yet another intriguing connection, between scattering amplitudes and the geometry of \emph{tropical Grassmannians} Tr$(k,n)$ \cite{Speyer2003}, or more accurately \emph{tropical configuration spaces} $\TC{k,n}$, inheriting the additional quotienting of Grassmannians mentioned before. This connection has first been established in the context of tree-level amplitudes in a generalised biadjoint scalar theory \cite{Cachazo2019,Cachazo:2019apa}. These arise as an extension of the Cachazo-He-Yuan formulation \cite{Cachazo:2013hca,Cachazo:2013iea} of the corresponing amplitudes in `usual' (cubic) biadjoint scalar theory as an integral over $\mathbb{P}^1$, to an integral over $\mathbb{P}^{k-1}$. Other aspects of these amplitudes have been studied more recently in \cite{Sepulveda:2019vrz,Borges:2019csl,Cachazo:2019ble,Early:2019zyi}.

Building on the geometric picture for amplitudes in the $k=2$ case \cite{Arkani-Hamed:2017mur}, it was elucidated in \cite{Drummond:2019qjk} that the (canonically ordered) generalised biadjoint scalar amplitude is equal to the volume of a region of $\TC{k,n}$, which we shall denote as the \emph{totally positive tropical configuration space}\footnote{In the literature, $\TPTC{k,n}$  is often also called totally positive tropical Grassmannian by abuse of notation.} $\TPTC{k,n}$ \cite{Speyer2005}. In \cite{Drummond:2019qjk}, it was also pointed out that cluster algebras provide triangulations of $\TPTC{k,n}$, 
and that the nature of infinities of the former can be interpreted as an infinitely redundant decomposition of the finite volume of the latter into smaller simplices. Therefore a possible way to cure the infinities is to prevent these redundant triangulations, which in essence picks out a particular finite subset of the variables of the cluster algebra.

In this paper, we test and explore the implications of this `tropical selection rule'. Given that redundant triangulations have been observed to occur also in finite cluster algebras, and preventing them leads to certain `beyond-cluster adjacency' restrictions on which letters can appear next to each other in the symbol, we first study $\TPTC{4,7}$, or equivalently the seven-particle amplitude. By comparing these beyond-cluster adjacency predictions with existing data on the amplitude through four loops \cite{Drummond:2014ffa,Dixon:2016nkn,Drummond:2018caf}, we establish that the NMHV amplitude does not satisfy them, namely $\TPTC{4,7}$ is `too small' to contain the NMHV helicity configuration. At the same time, the present data suggests that the MHV amplitude does satisfy the $\TPTC{4,7}$-adjacency, as well as certain additional beyond-cluster adjacency restrictions. In other words, we find that $\TPTC{4,7}$ is sufficient to describe the MHV amplitude, albeit, in a sense, `too big'.

Armed with this intuition, we then move on to study $\TPTC{4,8}$ and its tropical selection rule. In this manner, we obtain a finite set of 356 cluster ($\A$-)variables, which is expected to contain the rational symbol letters of the eight-particle MHV amplitude. The complete alphabet is also expected to contain square roots of the cluster variables, since the latter are also contained in the four-mass one-loop box (see for example \cite{Bourjaily:2013mma}), which starts contributing to the amplitude for $n\ge 8$. While these cannot appear in the cluster algebra per se, using a simple example of a rank-two affine cluster algebra, we show that it is possible to obtain square roots of very similar type, as a limit of an infinite mutation sequence. We find this example very relevant, given that the rank-two algebra considered is in fact a subalgebra of the Gr$(4,8)$ cluster algebra.

In our view, these results provide evidence that the relation between cluster algebras and tropical geometry may offer great promise for unravelling the analytic structure of $\N=4$ amplitudes, even though further work will be needed to flesh out the details. Before concluding, we also revisit the finite Gr$(3,8)$ cluster algebra, and show how to obtain the associated $\TPTC{3,8}$ generalised biadjoint scalar amplitude in a form containing a near-minimal amount of spurious poles.

This paper is organised as follows. We first review some general notions of tropical geometry and the construction of the Grassmannian and configuration space as well as their tropical versions in section \ref{sec:tropicalGeometry}. Following this, we give a quick overview of cluster algebras in section \ref{sec:clusterAlgebras} and review their connection to the totally positive tropical configuration space. Section \ref{sec:tptg47} contains our analysis of $\TPTC{4,7}$ and its implications for the seven-particle amplitude, whereas in section \ref{sec:tptg48} we focus on $\TPTC{4,8}$ and present its predictions for the rational symbol letters of the eight-particle alphabet.

Then, in section \ref{sec:clusterSeq} we show how square roots can arise as limits of cluster algebras. Section \ref{sec:tptg38} deals with $\TPTC{3,8}$ and the more compact representation we can obtain for the corresponding generalised biadjoint scalar amplitude, and finally section \ref{sec:Conclusions} contains our conclusions and outlook. Two appendices supplement the main text, reviewing the web-parameterisation of the $\TPTC{k,n}$ \cite{Speyer2005}, as well as the general construction of cluster algebras with coefficients \cite{CAIV}. 

Results similar to the ones presented in this paper, have been also independently obtained in \cite{Drummond:2019cxm,Arkani-Hamed:2019rds}.

\section{Tropical geometry and Grassmannians}
\label{sec:tropicalGeometry}

Tropical geometry is a relatively new field in mathematics, whose cornerstone is the mapping of polynomials, used to define algebraic varieties, to piecewise linear functions, defining in turn geometric objects known as polyhedral complexes. In this manner, it sometimes allows to answer questions about the former, in the much simpler context of the latter.
%Tropical geometry is a relatively new field in mathematics, in which a polyhedral complex is associated to an algebraic variety, thus sometimes allowing to answer questions about the latter in the much simpler context of the first. This is done by replacing addition with taking the minimum and multiplication by addition in a set of polynomials such that they become piecewise linear functions. 
There are many good introductory texts \cite{Katz2017} as well as reviews on the subject \cite{Mikhalkin2006,Maclagan2012,Brugalle2015}. 

Here, we will briefly review the basic concepts following \cite{Speyer2003,Speyer2005,Drummond:2019qjk} with a focus on introducing the totally positive tropical configuration space. The relation of its building blocks, known as rays, to the variables of Grassmannian cluster algebras, which will be discussed in the next section, will lie at the core of our analysis. We will use italic text to emphasize the main concepts and definitions, and also provide several examples to illustrate them,\footnote{So as to aid the reader, the beginning and end of each example along the text, will be denoted by boldface font and the \qedsymbol\, symbol, respectively.} as it is only relatively recently that they have been discovered by the physics community.

\subsection{Generalities on tropical geometry}
\label{sec:tropicalGeometry:generalities}
In essence, tropical geometry is the algebraic geometry over the tropical semifield $(\R\cup\left\{\infty\right\},\oplus,\otimes)$, which is defined as the set of real numbers with infinity on which addition is given by taking the minimum and multiplication by addition. The varieties that are of interest to us are usually defined over the complex numbers $\C$, such that we will only discuss this case, also known as tropical geometry with constant coefficients. Replacing this field by the tropical semifield allows us to construct the tropical variety as follows.

Given a polynomial $f \in \C\left[x_1,\dots,x_r\right]$ of the form $f=\sum_a c_ax_1^{m^a_1}\dots x_r^{m^a_r}$, where $c_a\in \C^*\equiv\C\setminus\left\{0\right\}$, we associate a tropical counterpart to it by replacing addition and multiplication as described above. We thus obtain, with the first sum to be understood as tropical addition over the terms, the tropical polynomial
\begin{equation}
	\label{equ:tropicalPolynomial}
	\trop f = {\sum_a}^\oplus \val{c_a}\otimes x_1^{\otimes m^a_1}\otimes\dots\otimes x_r^{\otimes m^a_r} = \min_a\left(\sum_{j=1}^r m^a_j\cdot x_j\right)\,,
\end{equation}
where $\text{val}$ denotes the valuation map, which in this case maps all elements of $\C^*$ to zero and zero to infinity. Similar to usual polynomials, we refer to the arguments of the minimum, that is terms of the form
\begin{equation}
	\sum_{j=1}^r m^a_j\cdot x_j\,,
\end{equation}
as the monomials of the tropical polynomial.

\begin{example}
	We demonstrate this and the following constructions on an example. Consider for this the polynomial $g(x_1,x_2) = 2x_1^3 + x_1^{-1}x_2 - x_2^2$. The three constant coefficients, which are $2$, $1$ and $-1$, all vanish under the valuation. We thus obtain the tropical polynomial
	\begin{equation}
		\trop g = \min\left(3x_1,-x_1+x_2,2x_2\right)\,.\qed
	\end{equation}
\end{example}

As can be seen from \eqref{equ:tropicalPolynomial}, the tropical polynomial $\trop f$ is a piecewise linear function from $\R^r$ to $\R$. This function is linear and differentiable everywhere except where the minimum is attained by at least two tropical monomials, which is where the function passes between regions of linearity. The set of such points in $\R^r$, that is where the function is not regular, is defined as the \emph{tropical hypersurface} $V\left(\trop f\right)$ associated to the polynomial.

\begin{example}
	In our previous example, the tropical hypersurface associated to $g$ is given by the union of solutions to each of the equations
	\begin{align}
		3x_1 &= -x_1 + x_2 \leq 2x_2\,, \nonumber\\
		3x_1 &= 2x_2 \leq -x_1 + x_2\,, \\
		2x_2 &= -x_1+x_2 \leq 3x_1\,,\nonumber
	\end{align}
	which is where the tropical polynomial is non-linear. This region is given by $V\left(\trop g\right) = \left\{x_2=4x_1, x_1\geq0 \right\} \cup \left\{x_2=3/2 x_1, x_1\leq0 \right\} \cup \left\{x_2=-x_1, x_1\geq0 \right\}$ and illustrated in figure \ref{fig:tropicalExample}. Note that the tropical hypersurface in this case is the union of three hypersurfaces in $\R^2$. \qed
	\begin{figure}[ht]
		\centering
		\includegraphics[width=0.3\textwidth]{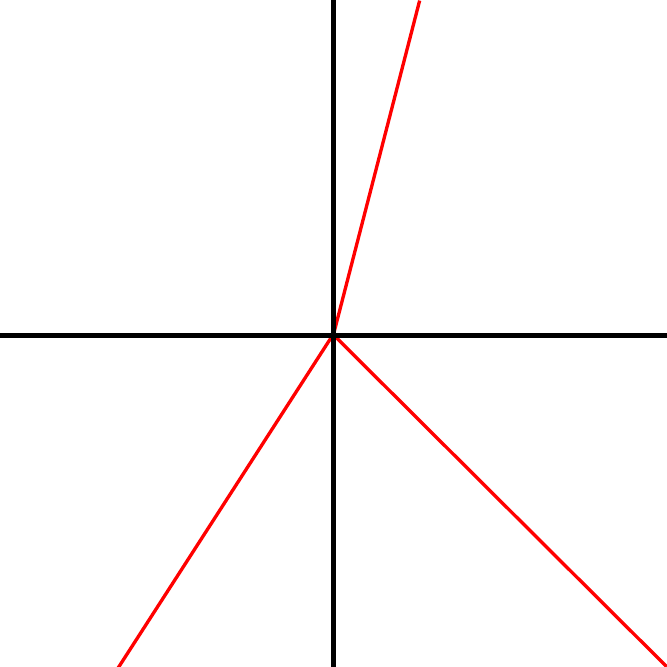}
		\caption{The tropical hypersurface associated to $g=2x_1^3 + x_1^{-1}x_2 - x_2^2$ is depicted in red.}
		\label{fig:tropicalExample}
	\end{figure}
\end{example}

In what follows, we will also need the notion of a tropical variety, which parallels the definition of an algebraic variety as the space of solutions to some polynomial equations. That is, for some ideal $I$ in $\C\left[x_1,\dots,x_r\right]$, the space of polynomials in $x_1,\dots,x_r$ with complex coefficients, the associated variety $V\left(I\right)$ is the space of points in $\C^r$ where all elements of the ideal vanish.\footnote{We remind the reader that an ideal is a subset of a ring closed under addition, such that any element of the ideal multiplied with any element of the ring again lands inside the ideal.} The associated tropical variety is then constructed by taking the intersection of the tropical hypersurfaces $V\left(\trop f\right)$ for all $f\in I$.

\subsection{Grassmannians and configuration spaces}
\label{sec:tropicalGeometry:grasConf}
Let us start by reviewing the definition of the Grassmannian before turning to its tropical version. The \emph{Grassmannian} $\G{k,n}$ is defined as the space of $k$-dimensional planes passing through the origin inside an $n$-dimensional space. Since a $k$-plane can be specified by a basis of $k$ vectors that span it, the Grassmannian may be equivalently described as the space of $k \times n$ matrices, modulo $GL(k)$ transformations that correspond to a change of basis.

%can be described as the space of $k$-dimensional linear subspaces of an $n$-dimensional vector space modulo the action of $GL(k)$, which leaves the subspaces invariant. 

For our purposes, a formulation of $\G{k,n}$ as the variety attached to a polynomial ideal, as reviewed in the previous section, will be more convenient. We thus consider the ring $\Z\left[p\right]$ of integer coefficient polynomials in the Plücker variables $\plAlt{i_1\dots i_k}$ for $1\leq i_1 < \dots < i_k \leq n$, which are nothing but the determinants made out of rows $i_1,\ldots, i_k$ of the aforementioned $n\times k$ matrix. From here on we denote the number of distinct Plücker variables of $\G{k,n}$ as
\begin{equation}
	D=\binom{n}{k}\,.
\end{equation}

Due to their nature as determinants, which are not all independent, Plücker variables will thus obey algebraic relations known as the Plücker relations,
\begin{equation}
	\label{equ:plueckerRelations}
	\plAlt{i_1\dots i_r [i_{r+1} \dots i_k}\plAlt{j_1 \dots j_{r+1}] j_{r+2} \dots j_k} = 0\,,
\end{equation}
where square brackets denote total antisymmetrisation with respect to the enclosed indices.

These homogeneous polynomials form the Plücker ideal $I_{k,n}$ in $\Z\left[p\right]$. The projective variety of this ideal, that is the set of zeros of these polynomials quotiented by global rescalings of the Plücker variables, $\plAlt{i_1 \dots i_k}\rightarrow t\cdot\plAlt{i_1 \dots i_k}$ with $t\in\C^*$, can be identified with the Grassmannian $\G{k,n}$. The dimension of this space is $k(n-k)$.

Note that the Plücker relations are not only invariant under global rescaling but also under the local scaling 
\be\label{eq:PluckerRescaling}
\plAlt{i_1 \dots i_k} \rightarrow t_{i_1} \dots t_{i_k}\plAlt{i_1 \dots i_k}\,,\,\, \text{for}\,\, t_1, \dots, t_n \in \C^*\,.
\ee
If we further quotient the Grassmannian by this local scaling, we obtain the \emph{configuration space} (of $n$ points in complex projective space $\mathbb{P}^{k-1}$),
\begin{equation}
	\Conf{k,n}\equiv \G{k,n}/\left(\C^*\right)^{n-1}= \text{Conf}_n\left(\mathbb{P}^{k-1}\right)\,.
\end{equation}
On the very right, we also include how the configuration space has been denoted in previous literature, however here we will prefer the notation on the very left to stress its relation to the Grassmannian. In what follows, we will denote the dimension of $\Conf{k,n}$ by $d$, where
\begin{equation}
	\label{eq:DimGrkn}
	d= (k-1)(n-k-1)\,.
\end{equation}

To construct the tropical version of the Grassmannian $\G{k,n}$, we follow the general procedure reviewed in section \ref{sec:tropicalGeometry:generalities} and first compute the tropical version of the polynomial generators of the Plücker ideal, 
\be
	I_{k,n}\to \text{Tr}I_{k,n}\,,\quad p_{i_1\ldots i_k}\to w_{i_1\ldots i_k}\,,
\ee
where we have also changed the notation for the tropical Plücker variables, so as to better distinguish them. We thus obtain a set of piecewise linear functions for each of which we compute its tropical hypersurface, given as those points where the minimum is attained by at least two tropical monomials.

The tropical analogue of the local rescaling \eqref{eq:PluckerRescaling},
\begin{equation}
	\label{equ:lineality}
	\plTrop{i_1 \dots i_k} \rightarrow a_{i_1} + \dots + a_{i_k} + \plTrop{i_1 \dots i_k}
\end{equation}
for any $a_1,\dots,a_n \in \R$, is known as lineality, and it similarly leaves the equations for the tropical hypersurfaces invariant. If we quotient by the global tropical scaling where $a_1=\ldots=a_k$, we obtain the \emph{tropical Grassmannian} $\TG{k,n}$, whereas quotienting by the local tropical scaling, we obtain the \emph{tropical configuration space}\footnote{Note that in both mathematics \cite{Speyer2003,Speyer2005} and physics literature \cite{Cachazo2019,Cachazo:2019apa,Drummond:2019qjk}, sometimes both $\TG{k,n}$ and $\TC{k,n}$ are referred to as the tropical Grassmannian, and are denoted as $\TG{k,n}$, by abuse of notation. The same terminology has also been used to denote the tropical variety obtained by tropicalising the Plücker relations, but without quotienting by any scaling of the type \eqref{equ:lineality}, neither local nor global.} $\TC{k,n}$.

In addition to the global and local scaling just considered, the tropical hypersurface conditions are also invariant under positive scalings of the tropical Plücker variables $\plTrop{i_1 \dots i_k}\rightarrow \lambda \plTrop{i_1 \dots i_k}$ for $\lambda\in\R^+$. This scaling follows from the homogeneity of the tropical monomials of the Plücker relations, which in turn is a consequence of the Plücker relations having constant $\C$ coefficients. 

In this sense, solutions to all tropical hypersurface equations form rays, that is a half-lines emanating from the origin of $\R^D$. The intersection of all these tropical hypersurfaces thus forms a \emph{polyhedral fan} in $\R^D$ \cite{Speyer2003}. This means that it is a set of \emph{cones} -- convex subregions of $\R^D$ containing the origin -- each of them obtained as the positive span of a given collection of the rays.

Depending on the number of linearly independent rays, these cones may have different dimensions, whereas the lower-dimensional cones arise as the boundaries of higher-dimensional ones. These faces of the fan are usually summarised in terms its $f$-\emph{vector} $f=(f_1,\dots,f_r)$, with $f_m$ being the number of $m$-dimensional faces. Here $r$ denotes the dimension of the fan, given by the highest dimensional cone. $\TG{k,n}$ is a polyhedral fan of $k(n-k)$, whereas $\TC{k,n}$, a polyhedral fan of dimension $(k-1)(n-k-1)$. Namely tropicalisation does not change the dimension.

\begin{example}
	We now demonstrate this general construction on the example of the tropicalisation of $\G{2,n}$. The Plücker ideals $I_{2,n}$ are generated by the polynomials
	\begin{equation}
		\label{equ:plueckerRelationExample}
		\plAlt{ij}\plAlt{ml} - \plAlt{im}\plAlt{jl} + \plAlt{il}\plAlt{jm}\,, \quad 1\leq i < j < m < l \leq n\,.
	\end{equation}

	Replacing addition with taking the minimum and multiplication with addition, we obtain the tropicalised Plücker relations given by
	\begin{equation}
		\label{equ:tropicalPolynomialExample}
		\min\left(\plTrop{ij}+\plTrop{ml},\, \plTrop{im}+\plTrop{jl},\, \plTrop{il}+\plTrop{jm}\right), \quad 1\leq i < j < m < l \leq n.
	\end{equation}
	For any of these polynomials, the tropical hypersurface is given as the set of points in $\R^{n(n-1)/2}$ solving the system of three equations
	\begin{align}
	\plTrop{ij}+\plTrop{ml} &= \plTrop{im}+\plTrop{jl}\leq\plTrop{il}+\plTrop{jm}\,, \nonumber \\
	\text{or}\quad\plTrop{ij}+\plTrop{ml} &= \plTrop{il}+\plTrop{jm}\leq\plTrop{ij}+\plTrop{ml}\,, \label{equ:tropicalHypersurfaceEx} \\
	\text{or}\quad\plTrop{il}+\plTrop{jm} &= \plTrop{im}+\plTrop{jl}\leq\plTrop{ij}+\plTrop{ml}\,. \nonumber
	\end{align}
	
	Now restricting to $\G{2,5}$, the tropical Plücker relations are given by equation \eqref{equ:tropicalPolynomialExample} with possible $\left\{i,j,m,l\right\}$ index combinations given by $\left\{1,2,3,4\right\}$, $\left\{1,2,3,5\right\}$, $\left\{1,2,4,5\right\}$, $\left\{1,3,4,5\right\}$ and $\left\{2,3,4,5\right\}$. 
	
	For each of these we obtain the associated tropical hypersurfaces by equations of the form \eqref{equ:tropicalHypersurfaceEx}. To obtain the points of the tropical Grassmannian, we have to solve all these five sets of equations simultaneously. Due to the invariance of the tropical Plücker variables under positive scalings, these solutions form rays in $\R^{10}$. \qed
\end{example}

\subsection{Totally positive tropical configuration space}
As the previously defined valuation maps all complex constants to zero, the tropical Plücker polynomials lose the information of the sign of the quadratic terms. For example, all three terms in \eqref{equ:plueckerRelationExample} end up with the same sign ($+1$) in the tropical polynomial \eqref{equ:tropicalPolynomialExample}. Restoring this information leads to the notion of the totally positive part of the tropical configuration space, denoted by $\TPTC{k,n}$. Following \cite{Speyer2005}, we will now demonstrate how it can be constructed.

Note that while there is a very similar construction to obtain the totally positive tropical Grassmannian $\TPTG{k,n}$,\footnote{In fact, in \cite{Speyer2005} such a construction is presented for any affine variety over the ring of Puiseux series, of which the Grassmannian is a special case.} we will focus on the totally positive part of the tropical configuration space, on which the remainder of this paper is based. 

The rays of the fan of $\TC{k,n}$ are given as $1$-dimensional intersections of tropical hypersurfaces. Equivalently, they are the simultaneous solution to the corresponding equations of the type of \eqref{equ:tropicalHypersurfaceEx}. Each of these equations comes from setting two tropical monomials equal, which in turn originate from the quadratic terms in the Plücker relations. One way to define the totally positive tropical configuration space $\TPTC{k,n}$ is to restrict the equations for each tropical hypersurface to those that come from Plücker quadratics with opposite signs \cite{Speyer2005,Drummond:2019qjk}. In the example of \eqref{equ:tropicalHypersurfaceEx} this removes the equation in the middle. 

A different way to construct $\TPTC{k,n}$ is to use the interaction of positivity and parameterisation \cite{Speyer2005}. The totally positive part of the configuration space, denoted as $\PConf{k,n}$, is obtained by restricting all Plücker variables and the local scalings to real, positive values.

As follows from the Plücker relations, the Plücker variables are not independent. In fact, there are $d$ independent variables required to describe $\PConf{k,n}$. By the means of the \emph{web-parameterisation}, the details of which are explained in appendix \ref{sec:webParam}, we obtain a parameterisation of the Plücker variables in terms of the web-variables $x_1,\dots,x_d$.

We collect the parameterisations of the Plücker variables into the bijective parameterisation function
\begin{equation}
\Phi : \left(\R^+\right)^{d} \rightarrow \PConf{k,n}\,. 
\end{equation}
By construction this maps the $d$ web-variables to a vector in $\left(\R^+\right)^D$ quotiented by local scalings, eq.~\eqref{eq:PluckerRescaling}, whose components $\Phi_{i_1\dots i_k}$ are the lexicographically ordered, independent parameterised Plücker variables $\plAlt{i_1\dots i_k}\left(x_1,\dots,x_d\right)$.

A parameterisation of the totally positive tropical configuration space $\TPTC{k,n}$ can now be obtained by tropicalising this parameterisation
\begin{equation*}
	\begin{tikzpicture}[scale=1.6]
		\node at (0,0) (a) {$\Phi_{i_1\dots i_k}\left(x_1,\dots,x_d\right) = \plAlt{i_1\dots i_k}\left(x_1,\dots,x_d\right)$};
		\node at (0,-1) (b) {$\left(\trop\Phi\right)_{i_1\dots i_k}\left(\tilde{x}_1,\dots,\tilde{x}_d\right) = \plTrop{i_1\dots i_k}\left(\tilde{x}_1,\dots,\tilde{x}_d\right)\,,$};
		\draw[->](a) edge node[right] {$\trop$} (b);
	\end{tikzpicture}
\end{equation*}
whereas we label the tropical web-variables as $\tilde{x}_1,\dots,\tilde{x}_d$. We thus obtain for every tropical Plücker variable $\plTrop{i_1 \dots i_k}$ a tropical polynomial. Similar to before, collecting these parameterised tropical Plücker variables results in a parameterisation of the totally positive tropical configuration space \cite{Speyer2005}
\begin{equation}
	\label{equ:webParam}
	\trop\Phi : \R^{d} \rightarrow \TPTC{k,n}\,.
\end{equation}

This map not only is a parameterisation of $\TPTC{k,n}$ but also encodes its structure. Each tropical Plücker variable $\plTrop{i_1\dots i_k}$ is a tropical polynomial in $\tilde{x}_1,\dots\tilde{x}_d$. For each tropical polynomial, we compute the associated tropical hypersurface, which is a set of hypersurfaces in $\R^d$ dividing the space into the regions of linearity of the piecewise linear function. We thus obtain a complete fan\footnote{A $d$-dimensional fan is complete if the union of its cones covers the entire ambient space $\R^d$.} in $\R^d$ for every Plücker variable, whose top-dimensional cones are given by the regions of linearity of its tropical parameterisation. The \emph{simultaneous refinement}\footnote{The simultaneous refinement of a set of fans is another fan, of which every cone is contained in one cone of every fan. Pictorially, it can be imagined as overlaying all of the fans. This causes some of the cones to intersect, creating new rays. Furthermore, this splits up the cones into smaller segments, which are then taken as the cones of the new refined fan.} of all these fans results in the \emph{Speyer-Williams fan} $F_{k,n}$, the fan associated to $\TPTC{k,n}$. Its top-dimensional cones describe the regions of linearity of $\trop\Phi$.

The simultaneous refinement equivalently amounts to taking the union of the tropical hypersurfaces of all Plücker variables. These hypersurfaces together divide $\R^d$ into many chambers, the cones of the fan $F_{k,n}$, which are precisely those regions, where all of the tropical parameterisations of the Plücker variables are linear simultaneously. The $1$-dimensional intersections of the hypersurfaces are the rays of this fan.

\begin{example}
	Let us continue our previous example and construct the totally positive tropical configuration space $\TPTC{2,5}$. By using the web-diagrams, we obtain the following parameterisation of the Plücker variables
	\begin{align}
		\Phi\left(x_1,x_2\right) &= \left(\plAlt{12},\plAlt{13},\plAlt{14},\plAlt{15},\plAlt{23},\plAlt{24},\plAlt{25},\plAlt{34},\plAlt{35},\plAlt{45}\right)\left(x_1,x_2\right) \nonumber\\&= \left(1,1,1,1,1,1+x_1,1+x_1+x_1x_2,x_1,x_1+x_1x_2,x_1x_2\right).
	\end{align}
	It immediately follows that when restricting the web-variables $x_1,x_2$ to real, positive values, the Plücker variables are also real and positive. Furthermore, by direct computation we see that the Plücker variables in this parameterisation satisfy the Plücker relations such that this, upon identification of vectors related by positive local scaling, indeed parameterises the totally positive configuration space.

	To obtain the tropical version of the totally positive configuration space, we tropicalise the parameterisation and obtain
	\begin{align}
	\trop\Phi\left(\tilde{x}_1,\tilde{x}_2\right) &= \left(\plTrop{12},\plTrop{13},\plTrop{14},\plTrop{15},\plTrop{23},\plTrop{24},\plTrop{25},\plTrop{34},\plTrop{35},\plTrop{45}\right)\left(\tilde{x}_1,\tilde{x}_2\right) \nonumber\\
	&= \left(0,0,0,0,0,\min\left(0,\tilde{x}_1\right),\min\left(0,\tilde{x}_1,\tilde{x}_1+\tilde{x}_2\right),\tilde{x}_1,\min\left(\tilde{x}_1,\tilde{x}_1+\tilde{x}_2\right),\tilde{x}_1+\tilde{x}_2\right).
	\end{align}

	As before, we can associate a tropical hypersurface to each of the piecewise linear function in $\trop\Phi$ as the set of points where the minimum is attained at least twice. Considering for example the tropical parameterisation of $\plTrop{25}$, this leads to $V\left(w_{25}\right)=\left\{\tilde{x}_1 = 0, \tilde{x}_2 \geq 0 \right\} \cup \left\{\tilde{x}_2 = 0, \tilde{x}_1 \leq 0 \right\} \cup \left\{\tilde{x}_2 + \tilde{x}_1 = 0, \tilde{x}_2 \leq 0 \right\}$. These three hypersurfaces form a fan in $\R^{2}$ consisting of the three rays and cones depicted in figure \ref{fig:fanExample}.\qed
	\begin{figure}[ht]
		\centering
		\includegraphics[width=0.35\textwidth]{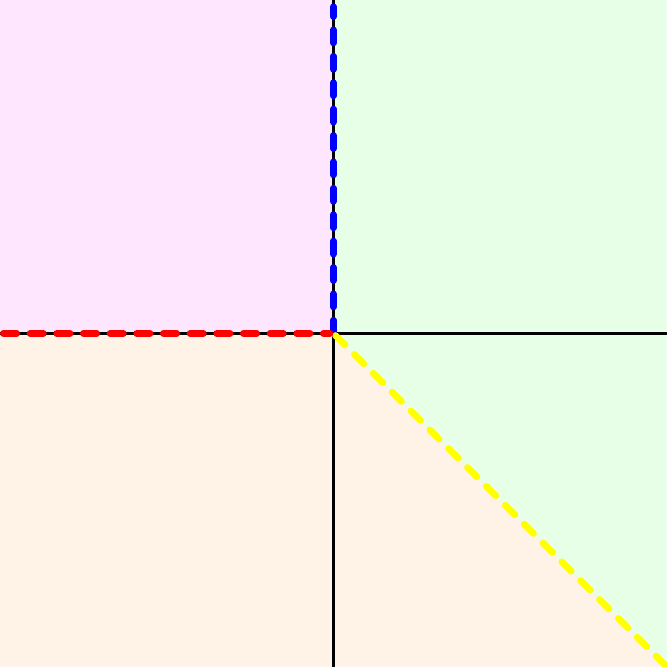}
		\caption{Fan associated to the tropicalised Plücker variable $\plTrop{25}$ of $\TPTC{2,5}$. The rays are depicted as dashed lines in blue, red and yellow, respectively. The cones are depicted in the composite color of the two rays by which they are spanned.}
		\label{fig:fanExample}
	\end{figure}
\end{example}

Note that in the construction of the totally positive tropical configuration space we tropicalised the parameterisations of all Plücker variables. We could, of course, also tropicalise only a subset of these variables thus obtaining a different polyhedral fan\footnote{Not tropicalising some Plücker variables amounts to the removal of the associated hypersurfaces from the fan. In this sense we obtain a smaller fan, of which the full totally positive tropical configuration space is a refinement.}. In this paper we will only focus on the full totally positive tropical configuration space, however, these different fans might also be of interest to scattering amplitudes.

Let us also make clear in how far we refer to two different objects here. On the one hand, we have the totally positive tropical configuration space $\TPTC{k,n}$, a polyhedral fan of dimension $d$ embedded into $\R^D$ quotiented by local scalings and part of the full tropical configuration space. On the other hand we have $F_{k,n}$, a polyhedral fan of dimension $d$ embedded into $\R^d$. Whereas the map $\trop\Phi$ gives a parameterisation of the full totally positive tropical configuration space, it furthermore maps the rays and cones of $F_{k,n}$ to those of $\TPTC{k,n}$. It is in this sense that $F_{k,n}$ captures the structure of $\TPTC{k,n}$.

\section{Cluster algebras and their fans}
\label{sec:clusterAlgebras}
Already in \cite{Speyer2005} it was pointed out that the totally positive tropical configuration space $\TPTC{k,n}$, which we reviewed in the previous section, is closely related to cluster algebras associated to Grassmannians $\G{k,n}$. Before elaborating on this relation in subsection~\ref{sec:clusterAlgebras:tropG}, we will first briefly review cluster algebras and establish our conventions.

There exist many introductory reviews and original results about cluster algebras, both purely from a mathematical point of view \cite{1021.16017,1054.17024,CAIII,CAIV,Fomin2016,Fomin2017} as well as with an emphasis on their application to scattering amplitudes \cite{Golden2014}. Here we will restrict to the basic concepts of cluster algebras associated to Grassmannians. 

Note that in the previous chapter we denoted the Plücker variables as $\plAlt{i_1 \dots i_k}$ as is usual in the mathematics literature. However, as is more usual in the discussion of cluster algebras associated to Grassmannians, especially in their applications to scattering amplitudes, we will from now on denote the Plücker variables as $\pl{i_1 \dots i_k}$.

\subsection{Generalities on cluster algebras}
As was demonstrated in \cite{Scott2006}, the coordinate ring of the Grassmannian $\G{k,n}$ naturally carries the structure of a \emph{cluster algebra} of rank $d=(k-1)(n-k-1)$. This means that the coordinate ring is generated by a distinguished set of generators, the cluster $\A$-variables, which are organised in overlapping clusters. Each cluster consists of $d$ such variables $\vect{a}=\left(a_{1},\dots,a_{d}\right)$ as well as $n$ frozen variables $(a_{d+1},\dots,a_{d+n})$, the so-called coefficients. 

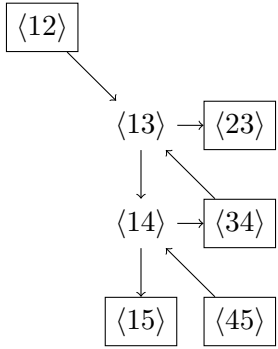
\begin{figure}[ht]
	\centering
	\begin{tikzpicture}[scale=1.3]
		\node at (-1,4) [rectangle,draw] (a) {$\pl{12}$};
		\node at (0,3) (b) {$\pl{13}$};
		\node at (0,2) (c) {$\pl{14}$};
		\node at (0,1) [rectangle,draw] (d) {$\pl{15}$};
		\node at (1,3) [rectangle,draw] (e) {$\pl{23}$};
		\node at (1,2) [rectangle,draw] (f) {$\pl{34}$};
		\node at (1,1) [rectangle,draw] (g) {$\pl{45}$};
		\draw[->](a)--(b) ;
		\draw[->](b)--(c);
		\draw[->](c)--(d);
		\draw[->](g)--(c);
		\draw[->](c)--(f);
		\draw[->](f)--(b);
		\draw[->](b)--(e);
	\end{tikzpicture}
	\caption{Quiver of the initial seed of the cluster algebra of $\G{2,5}$.}
	\label{quiv:gr25InitialQuiver}
\end{figure}

Together, these are arranged as nodes in a quiver whose adjacency properties are encoded in the antisymmetric integer adjacency matrix $B$. We denote the components of this matrix as $b_{ij}$. Note that frozen nodes are never connected. In figure \ref{quiv:gr25InitialQuiver} we show as an example the quiver of the initial seed of the $\G{2,5}$ cluster algebra.

The clusters within the cluster algebra are related by an operation called mutation. Starting from an initial seed $(\vect{a}_{0},B_{0})$ we can generate the entire algebra by mutating the unfrozen variables in each cluster. Mutating a cluster $(\vect{a},B)$ along the $j$-th variable, denoted as $\mu_j$, we obtain the new cluster $(\vect{a}',B')$. The adjacency matrix $B'$ is related to the previous one by
\begin{align}
	b'_{il} = 
		\begin{cases}
			-b_{il} \quad &\text{for}\,i=j\,\text{or}\,l=j\\
			b_{il} + \left[-b_{ij}\right]_+b_{jl} + b_{ij}\left[b_{jl}\right]_+\quad &\text{otherwise}
		\end{cases}\,,
\end{align}
whereas we use $\left[x\right]_+ = \max\left(0,x\right)$. The cluster variables $a_{i}$ are unchanged for $i\neq j$ and $a_{j}$ is mutated according to
\begin{equation}
	\label{equ:clusterMutation}
	a'_j = a_j^{-1}\left(\prod_{i=1}^{d+n}a_{i}^{\left[b_{ij}\right]_+} + \prod_{i=1}^{d+n}a_{i}^{\left[-b_{ij}\right]_+}\right).
\end{equation}
Note that the coefficients are called frozen variables as they are never mutated and do not change under mutation. To every unfrozen node and thus $\A$-variable in a cluster we further associate a cluster $\X$-variable, which is defined as
\begin{equation}
	\label{eq:xtoa}
	x_{i} = \prod_{l=1}^{d+n}a_{l}^{b_{li}}\,.
\end{equation}

As is well described in \cite{Golden2014}, we can associate a convex polytope to a cluster algebra \cite{Fomin2003,Chapoton2002}. This polytope is constructed by associating to each cluster a vertex, which are connected by lines corresponding to the mutations of the $\A$-variables. In this sense we obtain a $d$-dimensional finite polytope from a finite rank-$d$ cluster algebra.

We obtain a rank $d-m$ cluster subalgebra by additionally freezing $m$ $\A$-variables that together appear in a cluster. Mutation is then restricted to the $d-m$ unfrozen variables. In the cluster polytope, this subalgebra is associated to a codimension-$m$ face, which can correspondingly be identified with the frozen variables. 

\begin{figure}[ht]
	\centering
	\includegraphics[width=0.3\textwidth]{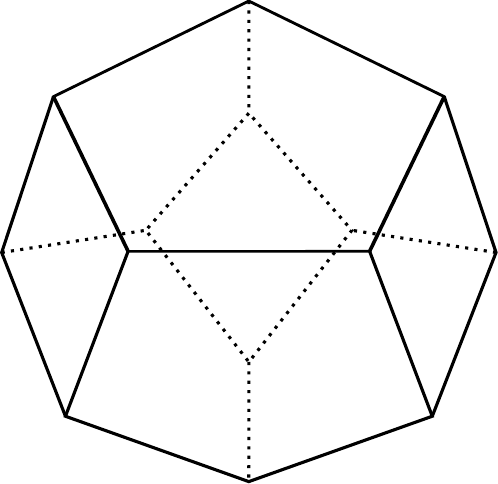}
	\caption{Cluster polytope of the cluster algebra of $\G{2,6}$.}
	\label{fig:gr26clusterPolytope}
\end{figure}

For example, freezing all but one variable, we obtain a rank-$1$ cluster algebra with only two variables, related by mutation. This is precisely the line connecting two vertices. In figure \ref{fig:gr26clusterPolytope} the cluster polytope of the cluster algebra of $\G{2,6}$ is given as an example.

Before we can discuss the relation of cluster algebras to the totally positive tropical configuration space we need one further component. We thus also attach the coefficient matrix $C$ to any cluster. We define $C_0=\unit$ for the initial seed and its mutation for $\mu_j$ as
\begin{equation}
	c'_{il} = 
		\begin{cases}
		-c_{il} \quad &\text{for}\,i=j\,\text{or}\,l=j\\
		c_{il} - \left[c_{ij}\right]_+b_{jl} + c_{ij}\left[b_{jl}\right]_+\quad &\text{otherwise}
		\end{cases}\,.
\end{equation}

\subsection{Cluster algebras and tropical configuration spaces}
\label{sec:clusterAlgebras:tropG}
As was discussed in \cite{Drummond:2019qjk} we can identify the $\X$-variables of the initial seed of the cluster algebra with the web-coordinates of the totally positive tropical configuration space. Furthermore, we can use the cluster algebra of $\G{k,n}$ to construct a fan in $\R^d$ that is closely related to the fan of the totally positive configuration space.

For this we associate to each $\A$-variable in the cluster algebra a ray $\g \in \R^d$, which can also be obtained in terms of a mutation rule. In this way we get $d$ rays for each cluster in the algebra, which can be taken to form a cone. The \emph{cluster fan} is then defined as the fan consisting of these rays and cones.

To organize the rays of the $\A$-variables of a cluster, we introduce the ray matrix $G$, a real $d\times d$ matrix whose columns are given by the rays of the cluster. Similar to before, the components are denoted by $g_{ij}$. Further following \cite{Drummond:2019qjk} we define the ray matrix of the seed to be $G_0=\unit$, that is the rays of the $\A$-variables of the cluster seed are given by the standard basis vectors. 

When mutating a given cluster along the $j$-th variable, the ray matrix $G$ mutates in terms of the adjacency matrix of the cluster seed $B_0$, the adjacency matrix $B$ and the coefficient matrix $C$ of the current cluster according to
\begin{equation}
	\label{equ:rayMutationRule}
	g'_{il} = 
		\begin{cases}
			g_{il} \quad &\text{for}\,l\neq j \\
			- g_{il} + \sum_{m=1}^d\left(g_{im}\left[-b_{mj}\right]_+ + b^0_{im}\left[c_{mj}\right]_+\right)  \quad &\text{for}\,l = j 
		\end{cases}\,.
\end{equation}

Performing all possible mutations, we obtain all rays and cones of the cluster fan, which is a polyhedral fan in $\R^d$. The cluster fan seems to have very special properties. All top-dimensional cones are simplicial, that is the $d$ rays of any cluster are linearly independent, and all codimension-$1$ cones are shared between precisely two top-dimensional cones. This latter property can be seen directly from the relation of this fan to the cluster polytope.

The cluster fan is equivalent to the dual of the cluster polytope. In this sense, a dimension-$m$ face of the fan corresponds to a codimension-$m$ face of the cluster polytope. For example, the codimension-$1$ faces of the fan correspond to the dimension-$1$ lines connecting precisely two clusters in the cluster polytope. This relation is further shown in table \ref{tab:clusterFanRelations}. 
\begin{table}[ht]
	\centering
	\begin{tabular}{|c|c|c|c|c|}
		\hline \hline
		\multirow{2}{*}{Algebra} & \multicolumn{2}{|c|}{Polytope} & \multicolumn{2}{|c|}{Fan} \\ \cline{2-5}
 		& Dim. & Type & Dim. & Type \\ \hline \hline
		Cluster & $0$ & Vertex & $d$ & Cone  \\ \hline
		Mutation & $1$ & Line & $d-1$ & Facet \\ \hline
		\multicolumn{1}{;{3pt/2pt}c;{3pt/2pt}}{$\vdots$} & \multicolumn{2}{;{3pt/2pt}c;{3pt/2pt}}{$\vdots$} & \multicolumn{2}{;{3pt/2pt}c;{3pt/2pt}}{$\vdots$} \\ \hline 
		$\A$-variable & $d-1$ & Facet & $1$ & Ray \\ \hline 
		\hline
	\end{tabular}
	\caption{Comparison of the faces of a cluster algebra of rank $d$, its polytope and the cluster fan.}
	\label{tab:clusterFanRelations}
\end{table}

As introduced in section \ref{sec:tropicalGeometry:grasConf} we express the number of dimension-$m$ faces of the fan in terms of its $f$-vector. With the equivalence of the cluster fan to the dual cluster polytope, the $m$-th component of the $f$-vector thus also captures the number of codimension-$m$ faces of the cluster polytope.

However, the most remarkable feature of the cluster fan is that it is a refinement of the fan $F_{k,n}$ of the totally positive tropical configuration space $\TPTC{k,n}$, as was first demonstrated by general techniques for $\TPTC{3,7}$ in \cite{Speyer2005} and for many other cluster algebras in \cite{Drummond:2019qjk}. With the cluster fan being simplicial, this implies that it triangulates $F_{k,n}$ and thus the totally positive tropical configuration space. It is this relation that allows computations of $\TPTC{k,n}$ by the simple algebraic operations on the cluster algebra of $\G{k,n}$.

\begin{figure}[ht]
	\centering
	\includegraphics[width=0.2\textwidth]{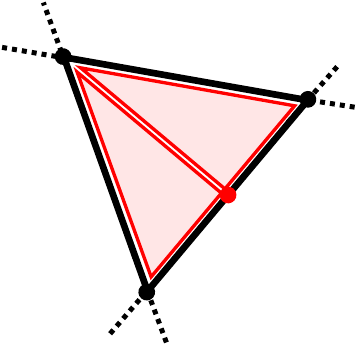}
	\caption{Illustrative example for redundant rays and triangulations. We draw the intersection of a simplicial cone of a $3$-dimensional fan with the unit sphere $S^2$ in black and that of two simplicial cones of another fan in red. The black lines correspond to the tropical hypersurfaces with rays of the associated fans at their intersections. The cone is further triangulated by two simplicial cones, drawn in red, due to the redundant ray that only lies on one of the hypersurfaces.}
	\label{fig:redundantRay}
\end{figure}

As can be seen for example in the case of $\G{3,8}$, which is discussed in section \ref{sec:tptg38}, in general the cluster algebra does not quite reproduce the rays of $F_{k,n}$. Some cones of the totally positive tropical configuration space are redundantly triangulated. This occurs whenever a ray associated to a cluster $\A$-variable is a \emph{redundant ray}, meaning that while it is a ray of the cluster fan, it is not a ray of $F_{k,n}$. This is illustrated in figure \ref{fig:redundantRay}. 

In general, a ray is redundant if the number of linearly independent tropical hypersurfaces it lies on, the \emph{ray rank}, is less then the maximal value $d-1$. In this case, the ray does not lie on a $1$-dimensional intersection of tropical hypersurfaces and is thus redundant.

The cones of the cluster fan can be fused along redundant faces to obtain the fan of the totally positive tropical configuration space. We consider a face of the cluster fan as a redundant face if it is not actually a face of $F_{k,n}$ but only apparent due to the triangulation of a non-simplicial cone by simplicial cones of the cluster algebra. Just as for redundant rays, we can determine whether a face is redundant or not by computing its face rank, the number of linearly independent tropical hypersurfaces on which the face lies. 

\begin{figure}[ht]
	\centering
	\includegraphics[width=0.22\textwidth]{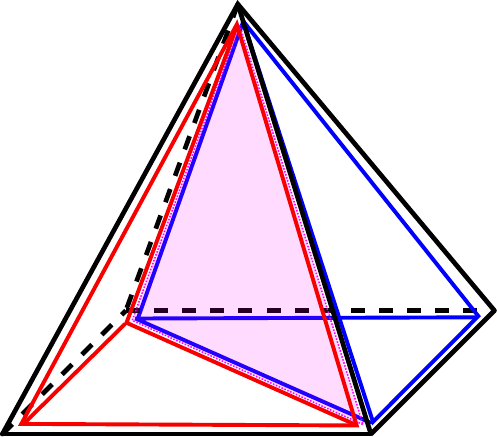}
	\caption{Illustrative example for fusing faces in triangulated cones. We draw the intersection a non-simplicial cone (the pyramid) of a $4$-dimensional fan with the unit sphere $S^3$. This cone is triangulated by two simplicial cones, drawn in red and blue. These two cones share a codimension-$1$ face along which we fuse. When doing so, the codimension-$2$ face on the base of pyramid is also fused and thus deleted.}
	\label{fig:fusingCones}
\end{figure}

A dimension-$m$ face is non-redundant if this number assumes the maximal value of $d-m$. In practice, for any $m$ we consider all dimension-$m$ faces and fuse them along their redundant dimension-$(m-1)$ boundaries, which are thus removed. This is illustrated in \ref{fig:fusingCones} for a pyramid with square base which is triangulated by two simplices.

\section{$\TG{4,7}$ and cluster adjacency of (N)MHV seven-particle amplitudes}
\label{sec:tptg47}
As we mentioned in the introduction, aside certain well-known rational factors, $n$-particle amplitudes in planar $\N=4$ SYM theory can be expressed in terms of multiple polylogarithms whose symbol alphabet consists of $Gr(4,n)$ cluster $\A$-variables, for $n=6,7$. Particularly for $n=7$, which will be the focus of this section, this alphabet  consists of 42 letters, which may be chosen as~\cite{Drummond:2014ffa}
\begin{align}
a_{11} &\propto {\pl{2367}}\,,
&
a_{41} &\propto
{\pl{2457}} \,,
\nonumber
\\
a_{21} &\propto
 {\pl{2567}}\,,
&
a_{51} &\propto
{\pl{1245} \pl{1367} - \pl{1267} \pl{1345}}\,,
\\
a_{31} &\propto
 {\pl{2347}}\,,
&
a_{61} &\propto
\pl{1356} \pl{1472} - \pl{1372} \pl{1456}
\,,
\nonumber
\end{align}
(up to proportionality factors made out of powers of the frozen variables $\pl{k\, k+1\, k+2\, k+4}$ which will not be relevant for our discussion) together with $a_{ij}$ obtained from $a_{i1}$ by cyclically permuting all integer indices $m \to ({m + j - 1})$ of the Pl\"ucker variables,\footnote{This is equivalent to cyclically permuting the columns of the $4 \times n$ matrix describing the $\Conf{4,n}$ space of kinematics, or in other words the momentum twistor variables~\cite{Hodges:2009hk}.} where the identification $m+n\sim m$ (here $n=7$) is implied.

In addition to the cyclic transformations we just described, the above alphabet is also invariant under the flip or reflection $m \to {n + 1 - m}$  of the Pl\"ucker indices, and together the two types of discrete transformations form the \emph{dihedral group}. After employing the supersymmetry of the theory, amplitudes with different external states can be combined to form a superamplitude, and in the planar limit the latter can also be shown to be dihedrally symmetric for any number of external particles, see for example \cite{Elvang:2009wd}. 

We should note however, that the dihedral symmetry of the superamplitude does not necessarily imply that its transcendental part, and thus the corresponding symbol alphabet, should inherit this property. This is indeed true for the  MHV superamplitude, whose gluon component corresponds to the configuration where all but two gluons have positive helicity; it is a consequence of the fact that it contains a single Nair-Parke-Taylor rational factor \cite{Nair:1988bq}, and thus both the latter and its transcendental coefficient must individually respect dihedral symmetry. For the NMHV case however (where all but three gluons have positive helicity) and beyond, there exist more than one independent rational factors and associated transcendental coefficients. In general these cannot be chosen to respect the dihedral symmetry separately, even though the entire superamplitude does.

Coming back to cluster algebras, more recently it has been observed~\cite{Drummond:2017ssj,Drummond:2018dfd} that they not only predict the symbol letters of $\N=4$ SYM (super)amplitudes, but also which letters are allowed to appear in two consecutive entries of the symbol. This is the so-called cluster adjacency property, which states that two letters can appear consecutively in the symbol, if and only if there exists a cluster containing both of them.

For our current case of interest, the seven-particle amplitude, cluster adjacency implies that only 840 out of the total $42\times 41=1722$ ordered pairs of letters can appear next to each other. In the aforementioned reference, it was checked that all then known seven-particle property were consistent with this restriction, but in fact a subset of $784$ out of 840 adjacencies occurred. In particular, the following cluster-adjacent pairs\footnote{Note that the sets~\eqref{equ:extendedClusterAdjacencyP1} and~\eqref{equ:extendedClusterAdjacencyP2} are related by a dihedral reflection.}
\begin{subequations}
	\begin{align}
		\label{equ:extendedClusterAdjacencyP1}
		(a_{21},a_{64})\quad&\text{\& cyclic}\\
		\label{equ:extendedClusterAdjacencyP2}	
		(a_{31},a_{65})\quad&\text{\& cyclic}\\
		\label{equ:extendedClusterAdjacencyP3}
		(a_{11},a_{41})\quad&\text{\& cyclic + parity}
	\end{align}
\end{subequations}
were seen to be missing from the amplitudes (although the pairs~\eqref{equ:extendedClusterAdjacencyP3} do appear in certain integrals contributing to the amplitude). We will denote these putative missing pairs as `beyond-cluster adjacency' restrictions. One of the main findings of this section is that these pairs do appear in the amplitude at higher loops, which will also have important implications for the geometry of the space of kinematics.

In particular, the statements of cluster adjacency have a geometric interpretation in terms of the cluster polytope. Given two variables that together appear in a cluster we can simultaneously freeze them, thus obtaining a cluster subalgebra whose rank compared to the full algebra is reduced by two. This subalgebra appears in the cluster polytope as a codimension-$2$ face. By cluster adjacency, the pairs of consecutive letters appearing in the symbol of an amplitude are hence restricted to the codimension-$2$ faces of the polytope of the associated cluster algebra. 

In this way, the $840$ allowed ordered pairs of the seven-particle amplitude are associated to the $399$ codimension-$2$ faces of the $\E_6$ cluster polytope, each corresponding to an unordered pair of distinct cluster variables, as well as the $42$ pairs of equal cluster variables. In this picture, the pairs further removed in beyond--cluster adjacency correspond to the removal of the corresponding codimension-$2$ faces from the cluster polytope.

\subsection{Cluster adjacency of $\TG{4,7}$}
As was first observed in \cite{Speyer2005}, the geometry of the dual to the $\E_6$ cluster algebra polytope is closely related to that of the totally positive configuration space $\TPTC{4,7}$. As was described in \cite{Drummond:2019qjk} and reviewed above, the origin of that is the triangulation of $F_{4,7}$ by the cluster fan, which is equivalent to the dual of the cluster polytope.

Starting from the initial cluster of the $\E_6$ cluster algebra, we can use the mutation rule eq.~\eqref{equ:rayMutationRule} to obtain the entire cluster algebra and hence the cluster fan. We find that in accordance with the literature its $f$-vector is given by $(42, 399, 1547, 2856, 2499, 833)$. Fusing the cones of the triangulating fan along redundant faces, we obtain a fan with $f$-vector $(42, 392, 1463, 2583, 2163, 693)$, which is precisely that of $F_{4,7}$ \cite{Speyer2005}. 

It can be checked that the interior of the obtained cones does not intersect any tropical hypersurfaces as well as that all faces actually lie on tropical hypersurfaces and together form a complete fan, implying that the cluster fan actually triangulates $F_{4,7}$. Furthermore, with all computations taking only on the order of minutes to complete, this highlights the efficiency of using cluster algebras as a computational tool for totally positive tropical configuration spaces.

The fusion of cones of the cluster fan translates to the fusion of vertices of the cluster polytope, which correspond to the clusters of the cluster algebra. As can be seen from the $f$-vectors, this results in the removal of seven codimension-$2$ faces from the polytope. The corresponding dimension-$2$ faces of the cluster fan are each spanned by two rays. From the correspondence of the rays with the cluster variables we can thus read off which pairs of cluster variables are removed. The results are given in table \ref{tab:missingDim2Faces}. It turns out that the seven faces removed in $\TPTC{4,7}$ with respect to the $\E_6$ cluster polytope are precisely those given by eq.~\eqref{equ:extendedClusterAdjacencyP2}.
\begin{table}[ht]
	\centering
	\begin{tabular}{|c|c|c|}
		\hline \hline
		$-e_1$ & $e_2 + e_4 - e_6$ & $(a_{35}, a_{62})$\\ \hline
		$e_3$ & $-e_2 + e_5$ & $(a_{32}, a_{66})$\\ \hline
		$e_4$ & $-e_1 + e_3 + e_5 - e_6$ & $(a_{37}, a_{64})$\\ \hline
		$-e_6$ & $e_1 + e_2 - e_3 - e_5$ & $(a_{31}, a_{65})$\\ \hline
		$e_1 - e_2$ & $e_1 - e_2 - e_4 + e_6$ & $(a_{34}, a_{61})$\\ \hline
		$e_2 - e_3$ & $e_2 + e_4 - e_5 - e_6$ & $(a_{33}, a_{67})$\\ \hline
		$e_1 - e_4$ & $e_1 - e_3 - e_5$ & $(a_{36}, a_{63})$\\ \hline \hline
	\end{tabular}
	\caption{Dimension-$2$ faces that are missing in $F_{4,7}$ compared to the $\E_6$ cluster fan. In the first two columns, the two rays spanning the face are given. In the final column we state the cluster variables associated to these rays.}
	\label{tab:missingDim2Faces}
\end{table}

Note that only the faces related to eq.~\eqref{equ:extendedClusterAdjacencyP2} and not their reflected counterparts are missing in $\TPTC{4,7}$. This implies that the totally positive tropical configuration space is not symmetric under a dihedral reflection, which as we mentioned at the beginning of this section, is a symmetry of the cluster algebra and the associated superamplitudes. There are two ways to complete $F_{4,7}$ to a reflection symmetric fan -- either by including the missing faces or by further removing their reflection. 

Further removing faces results in a fan with $f$-vector $(42, 385, 1393, 2387, 1995, 651)$, where the dimension-$2$ faces missing with respect to the $\E_6$ cluster fan are precisely those given by eqs.~\eqref{equ:extendedClusterAdjacencyP1} and \eqref{equ:extendedClusterAdjacencyP2}. This fan was obtained by removing the reflection of the seven missing dimension-$2$ faces, which in turn causes higher-dimensional faces to fuse in a way reverse but equivalent to the fusing of cones described in section \ref{sec:clusterAlgebras:tropG}. 

\subsection{Cluster adjacency of seven-particle amplitudes}
As stated in \cite{Drummond:2017ssj}, general cluster adjacency was verified to be obeyed in all known MHV and NMHV seven- as well as nine-point amplitudes and is believed to apply to any BDS-like subtracted amplitude. Considering that the totally positive tropical configuration space might explain (part of) the beyond--cluster adjacency property that was observed in seven-point MHV amplitudes, one might wonder whether it serves as a more suited geometric object than the cluster polytope to describe some of the structure of these amplitudes. 

While it only captures the restriction of eq.~\eqref{equ:extendedClusterAdjacencyP2} and thus is still too large to properly describe MHV amplitudes, it should at least not be too small. We therefore analyzed the beyond--cluster adjacency properties of the known seven-particle MHV and NMHV amplitudes by looking at all consecutive pairs of letters appearing in their symbols. 

In \cite{Dixon:2016nkn,Drummond:2014ffa} the symbols of the seven-particle MHV amplitude up to four loops are given. We found that up to two loops the set of pairs of consecutive letters is smaller then dictated by the three restrictions of eqs.~\eqref{equ:extendedClusterAdjacencyP1}, \eqref{equ:extendedClusterAdjacencyP2} and \eqref{equ:extendedClusterAdjacencyP3}. Starting at three loops, however, the set of pairs of consecutive letters in these amplitudes appears to stabilize with a total number of $371$ pairs. These are all pairs allowed by cluster adjacency with the exception of those stated in all three restrictions.

In \cite{Drummond:2018caf,Dixon:2016nkn,CaronHuot:2011ky} the symbols of NMHV seven-particle amplitudes up to four loops are given. Up to three loops, we again find that the set of pairs of consecutive letters is smaller than dictated by beyond--cluster adjacency. At four loops, however, we find all $399$ pairs of consecutive letters in agreement with cluster adjacency, that is also the pairs of eqs.~\eqref{equ:extendedClusterAdjacencyP1} and \eqref{equ:extendedClusterAdjacencyP2} do appear. These results are also summarised in table \ref{tab:consecutivePairsInHeptagon}.
\begin{table}[ht]
	\centering
	\begin{tabular}{|l|c|c|c|c|}
		\hline \hline
		Loops & $1$ & $2$ & $3$ & $4$ \\ \hline \hline
		MHV & $42$ & $210$ & $371$ & $371$ \\ \hline
		NMHV & $63$ & $294$ & $343$ & $399$ \\ \hline 
		Cluster adjacency & \multicolumn{4}{c|}{$399$} \\
		+ eqs. \eqref{equ:extendedClusterAdjacencyP1} \& \eqref{equ:extendedClusterAdjacencyP2} & \multicolumn{4}{c|}{$385$} \\
		+ eqs. \eqref{equ:extendedClusterAdjacencyP3} & \multicolumn{4}{c|}{$371$} \\ \hline
		\hline
	\end{tabular}
	\caption{Number of unordered pairs of distinct consecutive letters in the symbols of the seven-particle amplitudes at given loop level. Note that all the found pairs obey general cluster adjacency, that is the letters appear together in a cluster of the $\E_6$ cluster algebra. In the last three rows we give the number of consecutive pairs of letters that are theoretically possible considering the seven-particle alphabet and imposing only cluster adjacency or also the restrictions to it.}
	\label{tab:consecutivePairsInHeptagon}
\end{table}

The analysis of the adjacent letters in the MHV and NMHV seven-particle amplitudes thus suggests that while the tropical configuration space $\TPTC{4,7}$ might explain the adjacency properties of MHV amplitudes, it is actually too large to fully describe them, at least up to loop four. Furthermore, the totally positive tropical configuration space certainly is too small to describe the adjacency properties of the NMHV amplitudes, whose pairs of adjacent letters fully saturate those obtained by imposing cluster adjacency only.

\subsection{Weight-2 words in seven-particle amplitudes}
Not every tensor is the symbol of some function. A necessary and sufficient condition for a tensor
\begin{equation}
	 \sum_{i_1,\dots,i_k} c_{i_1\dots i_k} \log a_{i_1}\otimes \dots \otimes \log a_{i_k}
\end{equation}
to be the symbol of some function is the integrability condition \cite{chen1977}. Considering weight-2 words, that is symbols consisting of two letters, this condition is given by
\begin{equation}	
	\sum_{i,j}c_{ij} d\log a_i\wedge d\log a_j = 0\,.
\end{equation}

We hence analyse the independent integrable weight-2 words that appear in the symbols of the MHV and NMHV seven-particle amplitudes. In theory, the weight-2 words could be formed out of all $42$ seven-particle letters. Being the symbols of amplitudes, integrability has to be imposed on these combinations. Further imposing cluster adjacency, we obtain $573$ independent integrable weight-2 words. If we also impose the beyond--cluster adjacency restrictions~\eqref{equ:extendedClusterAdjacencyP1} and \eqref{equ:extendedClusterAdjacencyP2}, this number is further reduced to $559$. These numbers are listed in table \ref{tab:independentWeight2IntegrableWords}.

Based on the same data as the analysis of cluster adjacency for seven-particle amplitudes, the number of independent integrable weight-2 words that appear in these amplitudes are displayed in table \ref{tab:independentWeight2IntegrableWords}. Similar to before, we find that while the weight-2 integrable words appearing in the MHV amplitudes up to four loops do not saturate the maximally possible number and are thus compatible with beyond--cluster adjacency, the NMHV amplitudes only follow cluster adjacency. At four loops, the restrictions of eqs.~\eqref{equ:extendedClusterAdjacencyP1} and \eqref{equ:extendedClusterAdjacencyP2} are too strict and do not hold for the NMHV seven-particle amplitude.
\begin{table}[ht]
	\centering
	\begin{tabular}{|l|c|c|c|c|}
		\hline \hline
		Loops & $1$ &$2$ & $3$ & $4$ \\ \hline \hline
		MHV & $1$ & $98$ & $489$ & $531$ \\ \hline
		NMHV & $15$ & $294$ & $496$ & $573$ \\ \hline 
		Integrability & \multicolumn{4}{c|}{$1035$} \\ 
		+ Cluster adjacency & \multicolumn{4}{c|}{$573$} \\
		+ eqs. \eqref{equ:extendedClusterAdjacencyP1} \& \eqref{equ:extendedClusterAdjacencyP2} & \multicolumn{4}{c|}{$559$} \\
		+ eqs. \eqref{equ:extendedClusterAdjacencyP3} & \multicolumn{4}{c|}{$545$} \\ \hline
		 \hline
	\end{tabular}
	\caption{Number of independent integrable weight-2 words appearing in the seven-particle amplitudes at given loop level. Note that the sets are inclusive with increasing loop number, that is the words appearing at lower loop are part of those at higher loop. In the last four rows we give the number of independent weight-2 words that are theoretically possible considering the seven-particle alphabet and imposing only integrability or also cluster adjacency or also the restrictions to it.}
	\label{tab:independentWeight2IntegrableWords}
\end{table}

As for the pairs of adjacent letters, this data suggests that the totally positive tropical configuration space is too small to properly describe the seven-particle NMHV amplitudes, as the integrable weight-2 words again saturate those obtained by imposing cluster adjacency only. However, up to four loops the integrable weight-2 words in the seven-particle MHV amplitude do not even saturate those obtained from beyond--cluster adjacency, even with all three restrictions imposed. To confirm whether MHV amplitudes indeed saturate this 545-dimensional space, it would be interesting to compute their 5-loop correction.

\section{$\TG{4,8}$ and the rational eight-particle alphabet}
\label{sec:tptg48}
\begin{figure}[ht]
	\centering
	\includegraphics[width=0.3\textwidth]{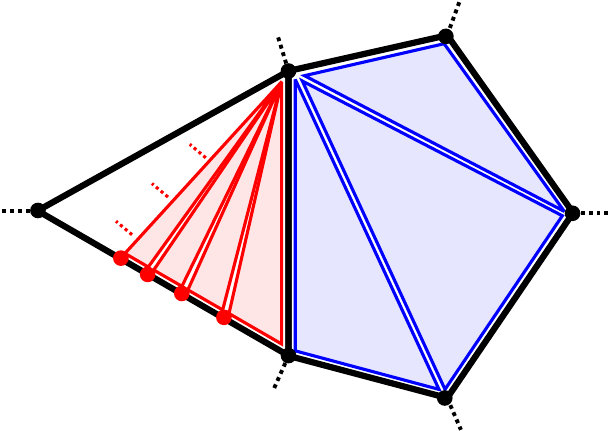}
	\caption{Illustrative example of an infinitely redundantly triangulated cone. We draw a simplicial cone of a $3$-dimensional fan intersected with the unit sphere $S^2$. The cones and the redundant rays from the infinitely redundant triangulation are drawn in red, the non-redundant triangulation in blue.}
	\label{fig:infiniteRedundantTriangulation}
\end{figure}

As is generally known, the cluster algebras of $\G{4,n}$ for $n\geq 8$ are infinite \cite{Scott2006}. Assuming that the cluster algebra always triangulates the totally positive tropical configuration space, the infinity of the cluster algebra implies that the cluster fan redundantly triangulates some cones of $F_{k,n}$ with infinitely many simplicial cones. This is illustrated in figure \ref{fig:infiniteRedundantTriangulation}. 

However, as proposed in \cite{Drummond:2019qjk} the totally positive tropical configuration space $\TPTC{4,n}$ provides a \emph{selection rule} to obtain a finite subset of these infinite cluster algebras. Whereas the cluster fan triangulates $F_{k,n}$, the cluster algebra sometimes produces redundant triangulations. These are triangulations that, using a redundant ray, divide a cone into more simplicial cones than actually needed. If the cluster algebra is truncated whenever a cluster with such redundant rays is encountered, mutation closes on a finite number of clusters.

In \cite{Drummond:2019qjk} it was stated that this \emph{truncated cluster algebra} contains $169192$ clusters with $356$ variables. Using the mutation rule \eqref{equ:rayMutationRule} we perform all possible mutations truncating whenever we obtain a cluster with at least one redundant ray. This results in a non-complete fan with $f$-vector \begin{equation}
f_{4,8} = \left(356,9408,90248,428988,1144532,1796936,1648184,817178,169192\right),
\end{equation}
thus confirming these results.

\subsection{Reducing Plücker variables}
As per the Laurent phenomenon, any cluster variable is a Laurent polynomial in the initial variables. While this parameterisation in terms of the independent Plücker variables has its advantages, for the applications to scattering amplitudes it is more practical to have the letters in the most compact form. Furthermore, from this perspective, it is perfectly fine if the letters depend on  all, also dependent, Plücker variables. 

The reader interested in the final form of the proposed rational eight-particle alphabet may just skip directly to the next section. In order to demonstrate the general algorithm presented here, we will make use of the example of a letter of degree two in the Plücker variables given by
\begin{equation}
	a = \frac{\pl{1245} \pl{1567} \pl{3456} + \pl{1256} \pl{1345} \pl{4567}}{\pl{1456}}\,.
\end{equation}

By construction, we can parameterise all cluster variables as rational functions in the variables of any cluster. For small cluster algebras, this is obtained by taking the cluster as the initial seed and performing all possible mutations. This, due to its size, is not practical for the (truncated) cluster algebra of $\G{4,8}$. However, once we have the alphabet in terms of some initial cluster, we can identify the degree one letters that correspond to the Plücker variables by comparing them in the unique web-parameterisation. 

This, for example, allows us to identify the following Plücker variable in the $\G{4,8}$ cluster algebra
\begin{equation}
	\frac{\pl{1245} \pl{1567} \pl{3456} + \pl{1256} \pl{1345} \pl{4567} + \pl{1235} \pl{1456} \pl{4567}}{\pl{1245} \pl{1456}} =\pl{3567}\,.
\end{equation}
Having identified a variable we can solve such rational expressions for any of the Plücker variables, allowing us to exchange any variable in the parameterisation, for example $\pl{1456}$ by $\pl{3567}$.

By cluster mutation any cluster variable is given as a rational function whose denominator is a monomial in the initial Plücker variables. To simplify this rational function, we thus change the parameterisation by exchanging any of the factors in the denominator with any of the other Plücker variables. In some cases, this causes the nominator to factor in such a way that parts of the denominator can be canceled. In our example, we therefore use the rational expression of the Plücker variables in terms of the initial ones to replace $\pl{1456}$. Using the expression for $\pl{3567}$ we obtain
\begin{equation}
	a = \frac{\left(\pl{1245} \pl{1567} \pl{3456} + \pl{1256} \pl{1345} \pl{4567}\right)\left(\pl{1245} \pl{3567} - \pl{1235} \pl{4567}\right)}{\pl{1245} \pl{1567} \pl{3456} + \pl{1256} \pl{1345} \pl{4567}}\,,
\end{equation}
such that the denominator cancels out completely and we obtain the letter in the fully simplified form of
\begin{equation}
	a = \pl{1245} \pl{3567} - \pl{1235} \pl{4567}\,.
\end{equation}

Whereas in this simple example one such replacement was sufficient, we can in general proceed in this way until there is no such change of parameterisation that simplifies the rational variable in the sense of reducing the number of factors in the denominator. If this does not result in the full denominator to cancel out, we attempt a different path of iterated changes of parameterisation. Note that in this way we effectively scan through all possible parameterisations. In fact, using this algorithm it was possible to eliminate the complete denominator of all cluster variables.

\subsection{Finite rational eight-particle alphabet}
\label{sec:tptg48:result}
In total, we find $356$ distinct unfrozen cluster $\A$-variables. Together with the $8$ frozen variables, we thus obtain $356$ dual conformally invariant rational letters. The full result is contained in the ancillary text file \texttt{OctagonAlphabet.m} attached to the \texttt{arXiv} submission of this article. All of the letters can be reduced to homogeneous polynomials in the Plücker variables and are, grouped by their degree, given by
\begin{itemize}
	\item $70$ letters of degree one, the distinct Plücker variables $\pl{ijkl}$,
	\item $120$ letters of degree two, $15$ quadratic generators with cyclic orbit size $8$,
	\item $132$ letters of degree three, $2$ cubic generators with $2$ cyclic images, $2$ cubic generators with $4$ cyclic images and $15$ cubic generators with $8$ cyclic images,
	\item $32$ letters of degree four, $4$ quartic generators with $8$ cyclic images and
	\item $10$ letters of degree five, $1$ quintic generator with $2$ and $1$ quintic generator with $8$ cyclic images.
\end{itemize}

For simplicity, we state the letters only modulo cyclic transformations, that is up to shifts of the Plücker indices $i\rightarrow (i + j) \mod 8$. Not considering the Plücker variables $\pl{ijkl}$ themselves, the alphabet is cyclically generated by $40$ distinct letters, which are summarised in tables \ref{tab:octagonAlphabetP1} and \ref{tab:octagonAlphabetP2}. Note that we only give one possible representation, which is related to other choices by the Plücker relations. However, using the web-parameterisation of the Plücker variables as defined above, we can always express all the cluster $\A$-variables in terms of the initial $x$-variables. In this parameterisation, all Plücker relations are solved therefore resulting in a unique representation.

For general cluster algebras, the Laurent phenomenon assures that any cluster variable can be written as a Laurent polynomial in terms of the initial variables. As can be seen from the results in the case of $\G{4,8}$, in the case of cluster algebras of Grassmannians there seems to be the much stronger statement that the variables are actual polynomials in the Plücker variables with $\pm 1$.

Even when taking into account that we have the freedom to add any Plücker relation, such that all integer coefficient may appear, the fact that we obtain actual polynomials with integer coefficients that also do have a representation with $\pm 1$ coefficients only is quite surprising. This is also present when passing to the web-parameterisation -- in contrast to the Plücker variables, the $x$-variables are well defined coordinates on the configuration space -- all cluster variables become polynomials with positive integer coefficients.

Note that this phenomenon is not restricted to the variables of the truncated cluster algebra but it also holds for all further letters of the full cluster algebra that were tested.

\begin{table}[ht]
	\centering
	\begin{tabular}{|c|c|c|}
		\hline \hline
		Degree & Cyclic Generator & \# \\ \hline \hline
		\multirow{10}{*}{$2$} 
		&$\pl{1457} \pl{2367} - \pl{1237} \pl{4567}$ & \\[2pt]
		&$\pl{1235} \pl{1467} - \pl{1234} \pl{1567}$ & \\[2pt]
		&$\pl{1245} \pl{3567} - \pl{1235} \pl{4567}$ & \multirow{6}{*}{$8$} \\[2pt]
		&$\pl{1256} \pl{4678} - \pl{1246} \pl{5678}, \quad \pl{1235} \pl{4678} - \pl{1234} \pl{5678}$ & \\ [2pt]
		&$\pl{1256} \pl{3467} - \pl{1267} \pl{3456}, \quad \pl{1256} \pl{1478} - \pl{1278} \pl{1456}$ & \\ [2pt]
		&$\pl{1245} \pl{3467} - \pl{1234} \pl{4567}, \quad \pl{1236} \pl{1478} - \pl{1234} \pl{1678}$ & \\ [2pt]
		&$\pl{1256} \pl{1347} - \pl{1234} \pl{1567}, \quad \pl{1256} \pl{3678} - \pl{1236} \pl{5678}$ & \\ [2pt]
		&$\pl{1245} \pl{2367} - \pl{1267} \pl{2345}, \quad \pl{1346} \pl{1578} - \pl{1345} \pl{1678}$ & \\ [2pt]
		&$\pl{1267} \pl{1358} - \pl{1235} \pl{1678}, \quad \pl{1347} \pl{2356} - \pl{1237} \pl{3456}$ & \\ [3pt]
		\hline
		\multirow{23}{*}{$3$} 
		&  $\pl{1236}\pl{1578}\pl{3457}-\pl{1237}\pl{1578}\pl{3456}-\pl{1235}\pl{1678}\pl{3457}$ & \multirow{2}{*}{$2$}\\[2pt]
		&$\color{highlightColour}\pl{1358}\pl{1367}\pl{2457}-\pl{1267}\pl{1358}\pl{3457}-\pl{1238}\pl{1567}\pl{3457}$ & \\[3pt]
		\cline{2-3}
		& $\pl{1258}\pl{1367}\pl{2456}-\pl{1238}\pl{1567}\pl{2456}-\pl{1258}\pl{1267}\pl{3456}$ & \multirow{2}{*}{$4$} \\[2pt]
		& $\pl{1236}\pl{1567}\pl{2458}-\pl{1238}\pl{1567}\pl{2456}-\pl{1236}\pl{1245}\pl{5678}$ & \\[3pt]
		\cline{2-3}
		& $\pl{1245}\pl{1567}\pl{2378}-\pl{1278}\pl{1567}\pl{2345}-\pl{1237}\pl{1245}\pl{5678}$ & \multirow{19}{*}{$8$} \\[2pt]
		& $\pl{1237}\pl{1568}\pl{3467}-\pl{1237}\pl{1678}\pl{3456}-\pl{1238}\pl{1567}\pl{3467}$ & \\[2pt]
		& $\pl{1237}\pl{1568}\pl{2467}-\pl{1237}\pl{1678}\pl{2456}-\pl{1238}\pl{1567}\pl{2467}$ & \\[2pt]
		& $\pl{1245}\pl{1568}\pl{3467}-\pl{1245}\pl{1678}\pl{3456}-\pl{1234}\pl{1568}\pl{4567}$ & \\[2pt]
		& $\pl{1256}\pl{1456}\pl{3478}-\pl{1256}\pl{1478}\pl{3456}-\pl{1234}\pl{1456}\pl{5678}$ & \\[2pt]
		& $\pl{1237}\pl{1458}\pl{2367}-\pl{1237}\pl{1678}\pl{2345}-\pl{1238}\pl{1457}\pl{2367}$ & \\[2pt]
		& $\pl{1256}\pl{1267}\pl{3478}-\pl{1256}\pl{1278}\pl{3467}-\pl{1234}\pl{1267}\pl{5678}$ & \\[2pt]
		& $\color{highlightColour} \pl{1246}\pl{1478}\pl{3567}-\pl{1278}\pl{1346}\pl{4567}-\pl{1236}\pl{1478}\pl{4567}$ & \\[2pt]
		& $\color{highlightColour} \pl{1246}\pl{1256}\pl{3478}-\pl{1246}\pl{1278}\pl{3456}-\pl{1234}\pl{1256}\pl{4678}$ & \\[2pt]
		& $\color{highlightColour} \pl{1456}\pl{2357}\pl{3678}-\pl{1678}\pl{2357}\pl{3456}-\pl{1235}\pl{3678}\pl{4567}$ & \\[2pt]
		& $\color{highlightColour} \pl{1358}\pl{1456}\pl{2367}-\pl{1238}\pl{1567}\pl{3456}-\pl{1236}\pl{1358}\pl{4567}$ & \\[3pt]
		& $\displaystyle \quad \pl{1235}\pl{1678}\pl{2345}-\pl{1238}\pl{1345}\pl{2567}$ & \\
		& $\displaystyle +\pl{1237}\pl{1345}\pl{2568}-\pl{1236}\pl{1578}\pl{2345}$ & \\[3pt]
		& $\displaystyle \quad \pl{1246}\pl{1356}\pl{2378}-\pl{1256}\pl{1378}\pl{2346}$ & \\
		& $\displaystyle -\pl{1236}\pl{1456}\pl{2378}-\pl{1236}\pl{1278}\pl{3456}$ & \\[3pt]
		& $\displaystyle \quad \pl{1246}\pl{1257}\pl{3458}-\pl{1245}\pl{1278}\pl{3456}$ & \\
		& $\displaystyle -\pl{1248}\pl{1256}\pl{3457}-\pl{1245}\pl{1267}\pl{3458}$ & \\[3pt]
		& $\color{highlightColour} \displaystyle \quad \pl{1256}\pl{3678}\pl{4578}-\pl{1236}\pl{4578}\pl{5678}$ & \\
		& $\color{highlightColour} \displaystyle +\pl{1678}\pl{2345}\pl{5678}-\pl{1345}\pl{2678}\pl{5678}$ & \\[3pt]
		\hline
		\hline
	\end{tabular}
	\caption{Cyclic generators of the eight-particle alphabet of degree $2$ and $3$. In the last column, the size of the cyclic orbit of each generator is given. The generators whose dihedral flip is not contained in the alphabet are highlighted in blue.}
	\label{tab:octagonAlphabetP1}
\end{table}
\begin{table}[ht]
	\centering
	\begin{tabular}{|c|c|c|}
		\hline \hline
		Degree & Cyclic Generator & \# \\ \hline \hline
		\multirow{13}{*}{$4$} 
			&$\color{highlightColour} \displaystyle \quad\pl{1245} \pl{1567} \pl{2378} \pl{3467}-\pl{1278} \pl{1567} \pl{2345} \pl{3467}$&\multirow{13}{*}{$8$}\\
			&$\color{highlightColour} \displaystyle	+\pl{1234} \pl{1237} \pl{4567} \pl{5678}-\pl{1237} \pl{1245} \pl{3467} \pl{5678}$ & \\
			&$\color{highlightColour} \displaystyle -\pl{1234} \pl{1567} \pl{2378} \pl{4567}$ &\\[6pt]
			&$\color{highlightColour} \displaystyle \quad\pl{1234} \pl{1567} \pl{1678} \pl{2345}-\pl{1267} \pl{1348} \pl{1567} \pl{2345}$&\\
			&$\color{highlightColour} \displaystyle +\pl{1267} \pl{1348} \pl{1457} \pl{2356}-\pl{1238} \pl{1267} \pl{1457} \pl{3456}$& \\
			&$\color{highlightColour} \displaystyle -\pl{1234} \pl{1457} \pl{1678} \pl{2356}$ &\\[6pt]
			&$\color{highlightColour} \displaystyle \quad\pl{1237} \pl{1458} \pl{1567} \pl{2368}-\pl{1238} \pl{1567} \pl{1678} \pl{2345}$&\\
			&$\color{highlightColour} \displaystyle +\pl{1236} \pl{1238} \pl{1457} \pl{5678}-\pl{1236} \pl{1237} \pl{1458} \pl{5678}$&\\
			&$\color{highlightColour} \displaystyle -\pl{1238} \pl{1457} \pl{1567} \pl{2368}$ &\\[6pt]
			&$\color{highlightColour} \displaystyle \quad\pl{1278} \pl{1678} \pl{2456} \pl{3456}-\pl{1278} \pl{1456} \pl{2678} \pl{3456}$&\\
			&$\color{highlightColour} \displaystyle +\pl{1256} \pl{1456} \pl{2678} \pl{3478}-\pl{1256} \pl{1678} \pl{2456} \pl{3478}$&\\
			&$\color{highlightColour} \displaystyle +\pl{1234} \pl{1678} \pl{2456} \pl{5678} -\pl{1246} \pl{1278} \pl{3456} \pl{5678}$& \\
			&$\color{highlightColour} \displaystyle-\pl{1234} \pl{1456} \pl{2678} \pl{5678}$ &\\[3pt]
		\hline
		\multirow{8}{*}{$5$} 
			&$\color{highlightColour} \displaystyle \quad\pl{1345} \pl{1458} \pl{1567} \pl{2367} \pl{2378} - \pl{1367} \pl{1458} \pl{1567} \pl{2345} \pl{2378}$&\multirow{4}{*}{$2$}\\
			&$\color{highlightColour} \displaystyle +\pl{1237} \pl{1238} \pl{1345} \pl{4567} \pl{5678} - \pl{1238} \pl{1345} \pl{1567} \pl{2378} \pl{4567}$&\\
			&$\color{highlightColour} \displaystyle +\pl{1237} \pl{1367} \pl{1458} \pl{2345} \pl{5678} - \pl{1235} \pl{1238} \pl{1678} \pl{3457} \pl{4567}$&\\
			&$\color{highlightColour} \displaystyle -\pl{1237} \pl{1345} \pl{1458} \pl{2367} \pl{5678}$&\\[3pt] 
		\cline{2-3}
			&$\color{highlightColour} \displaystyle \quad\pl{1235} \pl{1278} \pl{1678} \pl{2345} \pl{3456} - \pl{1235} \pl{1245} \pl{1678} \pl{2378} \pl{3456}$&\multirow{4}{*}{$8$}\\
			&$\color{highlightColour} \displaystyle +\pl{1235} \pl{1245} \pl{1568} \pl{2378} \pl{3467} - \pl{1235} \pl{1278} \pl{1568} \pl{2345} \pl{3467}$&\\
			&$\color{highlightColour} \displaystyle +\pl{1234} \pl{1278} \pl{1568} \pl{2345} \pl{3567} - \pl{1234} \pl{1245} \pl{1568} \pl{2378} \pl{3567}$&\\
			&$\color{highlightColour} \displaystyle +\pl{1234} \pl{1238} \pl{1245} \pl{3567} \pl{5678} - \pl{1235} \pl{1238} \pl{1245} \pl{3467} \pl{5678}$&\\[3pt]
		\hline 
		\hline
	\end{tabular}
	\caption{Cyclic generators of the eight-particle alphabet of degree $4$ and $5$. In the last column, the size of the cyclic orbit of each generator is given. The generators whose dihedral flip is not contained in the alphabet are highlighted in blue.}
	\label{tab:octagonAlphabetP2}
\end{table}

Similar to the reflection completion of $\TPTC{4,7}$ for the seven-particle amplitudes we analyse how the rational eight-particle alphabet behaves under dihedral symmetry. These transformations consist of cyclic shifts on the Plücker indices $i \rightarrow (i+j) \mod 8$ as well as dihedral flips $i\rightarrow(9-i)\mod 8$. We find that while the eight-particle alphabet is complete under the shift relations, it is not under dihedral flips. For example, flipping
\begin{equation}
	\pl{1358}\pl{1367}\pl{2457}-\pl{1267}\pl{1358}\pl{3457}-\pl{1238}\pl{1567}\pl{3457}
\end{equation}
we obtain a letter which is not included in the eight-particle alphabet obtained from the truncated cluster algebra, as can be seen by comparing them in the unique web-parameterisation. 

Again, there are two different ways to complete the alphabet under dihedral transformations. Including the dihedral image of the alphabet, we find a total of $440$ dual conformally invariant rational letters. Completing the alphabet by removing those letters whose dihedral images are not contained, we obtain $272$ dual conformally invariant rational letters. Note that all letters with degree four and five are not dihedrally complete in this sense. In the same way as before, we may construct these geometries and their $f$-vectors by deleting -- or adding -- the corresponding faces in $F_{4,7}$.

The $440$ rational letters we have obtained by taking the union of the $\A$-variables of the truncated cluster algebra of $\G{4,8}$ with their image under dihedral flips have been checked to be in agreement with those found by a different approach in \cite{Arkani-Hamed:2019rds}.

In analogy to the seven-particle case, where the intersection of the tropical fan with its image under reflections was sufficient to describe the MHV amplitude and its adjacency properties, it is reasonable to expect that the rational part of the eight-particle MHV alphabet will be similarly contained in the $272$ dihedrally complete letters mentioned above, and in fact may even be a subset thereof. Apart from the seven-particle expectations, this hypothesis is also the simplest that agrees with the available data on the 2-loop eight-particle symbol \cite{CaronHuot:2011ky} (which up to this order is fully rational).

In principle, this hypothesis could be tested by extending the amplitude bootstrap method mentioned in the introduction~\cite{Dixon:2011pw,Dixon:2011nj,Dixon:2013eka,Dixon:2014voa,Dixon:2014iba,Drummond:2014ffa,Dixon:2015iva,Caron-Huot:2016owq,Dixon:2016nkn,Drummond:2018caf,Caron-Huot:2019vjl,Caron-Huot:2019bsq}, so as to construct an ansatz for the 3-loop MHV 8-particle amplitude, and fix it completely by comparing with known data in the multi-Regge~\cite{Bartels:2008sc,Fadin:2011we,Bartels:2011ge,Lipatov:2012gk,Dixon:2012yy,Bartels:2013jna,Basso:2014pla,Drummond:2015jea,DelDuca:2016lad,DelDuca:2018hrv,Marzucca:2018ydt,DelDuca:2019tur} and collinear limits~\cite{Alday:2010ku,Basso:2013vsa,Basso:2013aha,Papathanasiou:2013uoa,Basso:2014koa,Papathanasiou:2014yva,Basso:2014nra,Belitsky:2014sla,Belitsky:2014lta}. It remains to be seen whether this task is also computationally feasible in practice, as the size of linear systems that need to be solved for the construction of the ansatz crucially depends on the number of letters, which is an order of magnitude larger than e.g. the seven-particle case.

\section{Cluster sequences and square roots}
\label{sec:clusterSeq}
In the previous section we discussed how to obtain the rational part of the eight-particle alphabet from the infinite cluster algebra associated to $\G{4,8}$. The rational letters, however, are not the complete story. Further to them, it is expected that also non-rational letters containing square roots appear in the symbols of eight-point NMHV amplitudes \cite{Prlina2018,Zhang:2019vnm}. These letters take the form of for example
\begin{equation}
	\label{eq:squareRootLetter}
	f_{il}f_{jm}\pm\left(f_{im}f_{jl}-f_{ij}f_{ml}\right)\pm\sqrt{\left(f_{ij}f_{ml}-f_{im}f_{jl}+f_{il}f_{jm}\right)^2 - 4f_{ij}f_{jm}f_{ml}f_{il}}\,,
\end{equation}
with $f_{ij} = \pl{ii+1jj+1}$. With the mutation being a rational transformation on the cluster variables, these square roots cannot appear after a finite number of mutations as the $\A$-variable in the cluster algebra itself.

However, there is a notion in which the infinity of the cluster algebra actually does produce square root variables. In \cite{Canakci2018} it was demonstrated that there is a sequence of mutations in a rank-2 cluster algebra of affine type, which is also infinite, such that a certain ratio of $\A$-variables converges to an algebraic letter of the form
\begin{equation}
	\label{equ:oldBetaLimit}
	\frac{a'_1+a_1 + \sqrt{\left(a'_1-a_1\right)^2+4}}{2 a_2}\,,
\end{equation}
with $a'_1 = (1+a_2^2)/a_1$. This result was obtained by identifying the sequence of $\A$-variables with a continued fraction. In this way, identities for these infinite continued fractions were used to obtain an equation for the limit.

Here we will generalise this result to also include frozen variables of certain type, but before doing so, let us first make some remarks on the classification of rank-2 cluster algebras. Whereas in the usual picture we associate to any cluster a quiver and its anti-symmetric adjacency matrix $B$, the most general perspective is to consider instead skew-symmetrisable adjacency matrices. Except for the notion of the cluster quiver, everything else in the construction of the cluster algebra -- including the mutation rules -- is unchanged.

Cluster algebras are usually classified in terms of the Dynkin classification \cite{1021.16017,1054.17024}. In the case of rank-$2$ cluster algebras, for any choice of $b$ and $c$ the matrix
\begin{equation}
	\label{equ:rank2AdjM}
	B_0 = \begin{pmatrix}
		0 & b \\
		-c & 0
	\end{pmatrix}
\end{equation}
is skew-symmetrisable and thus a valid adjacency matrix for the initial seed of a cluster algebra. The classification via Dynkin diagrams follows from the Cartan matrix associated to these adjacency matrices, which is given by
\begin{equation}
	A\left(B_0\right) = \begin{pmatrix}
	2 & -b \\
	-c & 2
	\end{pmatrix}\,.
\end{equation}

Depending on the determinant of this Cartan matrix, there are three different categories in the classification. For positive determinant, the Dynkin diagram is of finite type resulting for example in the usual $A_2$ cluster algebra. If the determinant is negative, the diagram is of hyperbolic type. Finally, there are two diagrams of affine type, given by $b = c = 2$ and $b=1$, $c=4$ and denoted by $A_1^{(1)}$ and $A_2^{(2)}$, respectively.

We will make use of the formulation of cluster algebras in terms of variables and coefficients, as reviewed in appendix \ref{sec:clusterAlgebras:coeffs}. In short this means that we work with the coefficients $y_i$ attached to each $\A$-variable of a cluster, which
are very similar to the $\X$-variables. Whereas the latter are monomials in the frozen and unfrozen variables obtained in terms of the adjacency matrix, the coefficients correspond to the part only involving the frozen variables and are related to the $\X$- and $\A$-variables by
\begin{equation}\label{eq:xtoay}
	x_{j} = \left(\prod_{i=1}^{d} a_{i}^{b_{ij}}\right)y_{j}\,.
\end{equation}
The coefficients also transform under mutations, and their transformation can be induced from the corresponding transformation of the $\A$-variables \eqref{equ:clusterMutation}, as well as their relation to the $\X$- and $y$- variables, eqs.~\eqref{eq:xtoa} and \eqref{eq:xtoay}. In the notations of appendix \ref{sec:clusterAlgebras:coeffs}, it is given directly in eq.~\eqref{equ:modClusterMutation2}.

Consider now the affine rank-$2$ cluster algebra of $A_{1}^{(1)}$ Dynkin type with \emph{principal coefficients}. In essence, this means that the initial seed has one frozen node connected to each of the unfrozen nodes, such that the frozen variables and coefficients are identified. This initial seed is thus given by $\left((a_1,a_2),(y_1,y_2),B_0\right)$ with the adjacency matrix given in the form of eq.~\eqref{equ:rank2AdjM} with $b=c=2$. The entire dynamics of the frozen variables is now encaptured in the coefficients $\left(y_1,y_2\right)$. These are related to the initial $\X$-variables by
\begin{equation}
	x_1 = a_2^{-2} y_1\,, \quad x_2 = a_1^2y_2\,.
\end{equation}

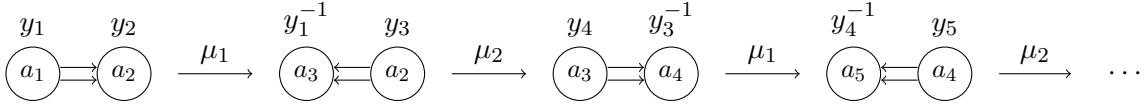
\begin{figure}[ht]
	\centering
	\begin{tikzpicture}[scale=1.2]
	\node[label=above:$y_1$,scale=0.9] at (0,0) [circle,draw] (a) {$a_1$};
	\node[label=above:$y_2$,scale=0.9] at (1,0) [circle,draw] (b) {$a_2$};
	
	\node[label=above:$y_1^{-1}$,scale=0.9] at (3,0) [circle,draw] (c) {$a_3$};
	\node[label=above:$y_3$,scale=0.9] at (4,0) [circle,draw] (d) {$a_2$};

	\node[label=above:$y_4$,scale=0.9] at (6,0) [circle,draw] (e) {$a_3$};
	\node[label=above:$y_3^{-1}$,scale=0.9] at (7,0) [circle,draw] (f) {$a_4$};	
	
	\node[label=above:$y_4^{-1}$,scale=0.9] at (9,0) [circle,draw] (g) {$a_5$};
	\node[label=above:$y_5$,scale=0.9] at (10,0) [circle,draw] (h) {$a_4$};
	
	\node at (12,0) (i) {$\cdots$};
	
	\draw[->] ([yshift=2pt]a.east) -- ([yshift=2pt]b.west);
	\draw[->] ([yshift=-2pt]a.east) -- ([yshift=-2pt]b.west);
	
	\draw[->,shorten >=10pt,,shorten <=10pt](b) edge node[above] {$\mu_1$} (c);

	\draw[->] ([yshift=2pt]d.west) -- ([yshift=2pt]c.east);
	\draw[->] ([yshift=-2pt]d.west) -- ([yshift=-2pt]c.east);

	\draw[->,shorten >=10pt,,shorten <=10pt](d) edge node[above] {$\mu_2$} (e);
	
	\draw[->] ([yshift=2pt]e.east) -- ([yshift=2pt]f.west);
	\draw[->] ([yshift=-2pt]e.east) -- ([yshift=-2pt]f.west);
	
	\draw[->,shorten >=10pt,,shorten <=10pt](f) edge node[above] {$\mu_1$} (g);
	
	\draw[->] ([yshift=2pt]h.west) -- ([yshift=2pt]g.east);
	\draw[->] ([yshift=-2pt]h.west) -- ([yshift=-2pt]g.east);
	
	\draw[->,shorten >=10pt,,shorten <=10pt](h) edge node[above] {$\mu_2$} (i);
	\end{tikzpicture}
	\caption{Cluster sequence $M$ in the affine rank-$2$ cluster algebra of $A_1^{(1)}$ Dynkin type.}
	\label{fig:mutSequ}
\end{figure}

In this cluster algebra we consider the mutation sequence $M: \mu_1\mu_2\mu_1\mu_2\dots$, which gives rise to a sequence of clusters with their $\A$-variables and coefficients, as depicted in figure \ref{fig:mutSequ}. We denote these sequences for $m=1,2,\dots$ by $a_m$ and $y_m$ with initial values $a_1,a_2$ and $y_1, y_2$, respectively.

We directly observe that the adjacency matrix is periodic in this sequence, whereas we always mutate on the node from which the two arrows originate. Not considering the first few elements, mutation takes $a_{m-1}$ to $a_{m+1}$, $y_{m-1}^{-1}$ to $y_{m+1}$ and $y_{m}$ to $y_m^{-1}$. Specialising the mutation rules \eqref{equ:modClusterMutation} and \eqref{equ:modClusterMutation2} to our case, we obtain the following recurrence relation for the sequences $a_m$ and $y_m$
\begin{align}
	y_{m+1}y_{m-1}\left(y_{m}\oplus 1\right)^2 &= y_m^2\quad \text{for}\,\,m\geq 4\,, \\
	a_{m+1}a_{m-1}\left(y_{m}\oplus 1\right) &= y_m + a_m^2\quad \text{for}\,\,m\geq 3\,,
\end{align}
with the initial values $y_4 = y_1^3y_2^2, y_3=y_1^2y_2$ and $a_3 = \left(y_1+a_2^2\right)/a_1$, and whereas the notation of the term in the parenthesis is also explained in appendix \ref{sec:clusterAlgebras:coeffs}. From the first few elements we can conjecture that $y_m = y_1^{m-1}y_2^{m-2}$ for $m\geq 4$ solves the recurrence relation for $y_m$. Indeed, this implies that $\left(y_m\oplus 1\right) = 1$ for all $m$ such that proving this inductively is a direct computation.

Numerically, one sees that the thus defined $a$-sequence diverges for $m\rightarrow\infty$. Therefore, we will instead consider the ratio
\begin{equation}
	\beta_m = \frac{a_m}{a_{m-1}}\,,
\end{equation}
which, as it turns out, does indeed converge. We hence rewrite the recurrence relation in terms of $\beta_m$. For this, we first make use of the separation principle by noting that for $m\geq 3$ we have $x_m = a_m^{-2}y_m$, such that the right hand side of the $a$-recurrence factors to $(1+x_m)a_m^2$. In this way, we obtain the modified recurrence relation
\begin{align}
	\beta_{m+1}\beta_{m-1} &= \left(1+x_m\right)\beta_m \quad \text{for}\,\,m\geq 3\,, \\
	x_{m+1}x_{m-1} &= \frac{x_m^2}{(1+x_m)^2}\quad \text{for}\,\, m\geq 4\,.
\end{align}

This factored form allows us to directly solve the recurrence relation for $\beta_m$ in terms of the $x$-sequence as
\begin{equation}
	\beta_m = \Gamma_m\left(1+x_1\right)\beta_2, \quad \Gamma_m = \prod_{j=3}^{m-1}\left(1+x_j\right)\,,
\end{equation}
such that it remains to obtain the limit of $\Gamma_m$ using the recurrence relation for $x_m$. To do so, we adapt the technique of \cite{Canakci2018} and translate the relations between the continued fractions used in there into the notation used here. After some generalisation, this leads to the relation for $\Gamma_m$ in terms of the initial $x_1$ and $x_2$ given by
\begin{equation}
	\left(1+x_1\right)\Gamma_m + \frac{x_1 x_2}{\left(1+x_1\right)\Gamma_{m-1}} = 1+x_1+x_1 x_2\,.
\end{equation}

Assuming that the limit $\Gamma=\lim_{m\rightarrow\infty}\Gamma_m$ exists, this relation gives us a quadratic equation for $\Gamma$. This equation has two solutions, one of which numerically matches the sequence considered here, such that we obtain the limit $\beta$ for the ratio $a_m/a_{m-1}$ as
\begin{equation}
	\label{equ:betaLimit}
	\beta = a_1^{-1}a_2\cdot \frac{1}{2}\left(1+x_1+x_1x_2 + \sqrt{\left(1+x_1+x_1x_2\right)^2-4x_1x_2}\right)\,.
\end{equation}

When taking $y_1=y_2=1$ we see that the $y$-sequence trivialises and the initial seed and cluster sequence are equal to that of the affine $A_1^{(1)}$ algebra without coefficients. In this case, this limit becomes equal to the previous result given in eq.~\eqref{equ:oldBetaLimit}.

The advantage of having computed the limit with principal coefficients is that we can now read off some interesting properties of this limit. Usually, any $\A$-variable of a cluster algebra with principal coefficients can be written as a monomial in the initial unfrozen variables times a polynomial in the initial $\X$-variables. From eq.~\eqref{equ:betaLimit} we see that a similar property holds for the limit of the ratio of $\A$-variables, which factors into such a monomial but now multiplied by an algebraic function.

Remarkably, this algebraic function is structurally very similar to the square-root letter of eq.~\eqref{eq:squareRootLetter}. Indeed, having obtained the limit for the affine rank-$2$ cluster algebra with principal coefficients is the first step towards the analysis of general coefficients, as is required for the application to the cluster algebra of $\G{4,8}$.

Furthermore, using such a factored form a vector is associated to every $\A$-variable, whose components are given by the exponents of the initial unfrozen $\A$-variables. These vectors are closely related to the rays of the totally positive tropical configuration space reviewed in section \ref{sec:clusterAlgebras:tropG} and the precise relation is discussed in appendix \ref{sec:clusterAlgebras:coeffs}. 

We observe that in this mutation sequence, the two rays associated to the $\A$-variables of every cluster both converge to $(-1,1)$, the vector similarly associated to the limit $\beta$. In this way, considering the limit of mutation sequences in $\G{4,8}$ might not only allow to derive the algebraic letter of eq.~\eqref{eq:squareRootLetter} but also its associated ray, which is required to obtain a full triangulation of $\TPTC{4,8}$.

\begin{figure}[ht]
	\centering
	\begin{tikzpicture}[scale=1.4]
	\node at (0,1) (6) {$a_1$};
	\node at (1,1) (9) {$a_2$};
	\node at (2,1) (2) {$a_3$};
	\node at (3.5,1) (5) {$a_4$};
	\node at (5,1) (1) {$a_6$};
	\node at (6,1) (7) {$a_8$};
	\node at (7,1) (8) {$a_9$};
	
	\node at (3.5,0) (3) {$a_5$};
	\node at (5,0) (4) {$a_7$};
	
	\draw[->] (6) -- (9);
	\draw[->] (9) -- (2);
	\draw[->] (2) -- (5);
	\draw[->] (1) -- (5);
	\draw[->] (7) -- (1);
	\draw[->] (8) -- (7);
	
	\draw[->] (3) -- (2);
	\draw[->] (3) -- (1);
	\draw[->] (4) -- (5);
	\draw[->] (3) -- (4);
	
	\draw[dashed,thick] (3.0,1.3) rectangle (4.0,-0.3);
	
	\draw[->] ([xshift=2pt]5.south) -- ([xshift=2pt]3.north);
	\draw[->] ([xshift=-2pt]5.south) -- ([xshift=-2pt]3.north);
	\end{tikzpicture}
	\caption{Principal part of a certain cluster in $\G{4,8}$. The affine $A_1^{(1)}$ cluster subalgebra is highlighted by the dashed rectangle.}
	\label{fig:gr48AffineA2Cluster}
	\end{figure}
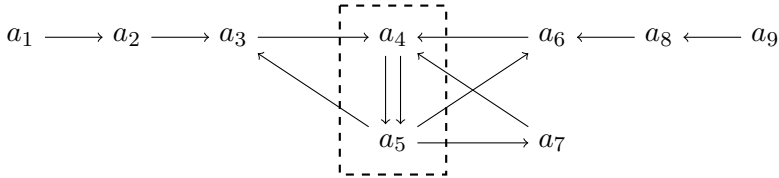

Indeed, the example considered here is closely related to the cluster algebra of $\G{4,8}$. To see this consider the cluster\footnote{The $\A$-variables of the cluster considered in figure \ref{fig:gr48AffineA2Cluster} are given by $a_1=\pl{3678}$, $a_2=\pl{1456}\pl{3678}-\pl{1678}\pl{3456}$, $a_3 = \pl{1347}\pl{2356}\pl{4678}-\pl{1237}\pl{3456}\pl{4678}-\pl{1347}\pl{2346}\pl{5678}$, $a_4=\pl{1367}$, $a_5=\pl{1347}\pl{2356}-\pl{1237}\pl{3456}$, $a_6=\pl{1267} \pl{1348} \pl{2356} - \pl{1234} \pl{1678} \pl{2356} - \pl{1238} \pl{1267} \pl{3456}$,  $a_7=\pl{1357}$,  $a_8=\pl{1267}\pl{1348}-\pl{1234}\pl{1678}$ and $a_9=\pl{1267}$.} depicted for simplicity without the frozen nodes in figure \ref{fig:gr48AffineA2Cluster}. The rays associated to the $\A$-variables of this cluster are all non-redundant, which is thus also a part of the truncated cluster algebra. Freezing all variables except those denoted in the figure as $a_4$ and $a_5$ results in an affine rank-$2$ cluster subalgebra of $A_1^{(1)}$ type with non-principal coefficients. Remarkably, mutating along $a_4$ results in an $\A$-variable with associated redundant ray. Therefore, it is plausible to assume that the same mutation sequence $M$ considered before leads to the square-root letters of the type eq.~\eqref{eq:squareRootLetter} as well as another non-redundant ray of $\TPTC{4,8}$.

With respect to the full triangulation of $\TPTC{4,8}$, it is worth mentioning that a different method to compactify the totally positive configuration space $\PConf{k,n}$ was reported in \cite{NimaAmplitudes2019}. In the case of finite cluster algebras, as for example those of $\G{4,7}$ and $\G{3,8}$, this leads to  geometries with the same $f$-vectors as that of the totally positive tropical configuration spaces computed by their cluster algebra triangulation in sections \ref{sec:tptg47} and \ref{sec:tptg38}, respectively. Moreover, for the infinite cluster algebra of $\G{4,8}$, this compactification similarly leads to a finite subset in this case containing $360$ letters.

Assuming that the announced compactification indeed reproduces the totally positive tropical configuration space, this suggests that compared to the truncated cluster algebra of $\G{4,8}$, the totally positive configuration space $\TPTC{4,8}$ contains $4$ more rays, presumably corresponding to algebraic letters that may be obtained via the limits of mutation sequences. However, with $18$ inequivalent algebraic letters being observed in the symbols of eight-particle NMHV amplitudes \cite{Zhang:2019vnm}, this suggests that similar to the seven-particle case the totally positive tropical configuration space is too small to describe these amplitudes.

Finally note that the results contained in this section, both the ratio of cluster variables as well as the ray, have also been obtained in \cite{Reading2018b} using the different perspective of scattering fans as introduced in \cite{Reading2018a}, see in particular corollary 3.11 and remark 3.14 in in the former reference.\footnote{To translate the notation used here to that of \cite{Reading2018b}, replace $x_i$ by $\hat{y}_i$ and $a_i$ by $x_i$. In this approach, the limit $\beta$ corresponds to the path-ordered product $\mathfrak{p}_\infty\left(a_1^{-1}a_2\right)$ within the $\vect{g}$-vector fan along the clockwise path from the limit ray $(-1,1)$ to the cone spanned by $(1,0)$ and $(0,1)$.}

\section{$\TG{3,8}$ and the generalised biadjoint scalar amplitude}
\label{sec:tptg38}
By the construction of a kinematic realisation of the associahedron, the amplitudes of biadjoint scalar $\phi^3$ theory can be related to volumes of geometric objects \cite{Arkani-Hamed:2017mur}. These so called kinematic associahedra are constructed as follows. Given $n$ light-like canonically ordered momenta $p_1,\dots,p_n$ we define the planar variables $X_{i,j}$ in terms of the Mandelstam variables $s_{i,i+1,\dots,j-1}$ as
\begin{equation}
	X_{i,j} = s_{i,i+1,\dots,j-1} = \left(p_i + p_{i+1} + \dots + p_{j-1}\right)^2\,,
\end{equation}
which allows us to expand the two-particle Mandelstam invariants $s_{i,j}$, which span the kinematic space $\mathcal{K}_n$, in terms of these planar variables as
\begin{equation}
	s_{i,j} = X_{i,j+1} + X_{i+1,j} - X_{i,j} - X_{i+1,j+1}\,.
\end{equation}
The kinematic associahedron is then defined as a region $\mathcal{A}_{n-3} \subset \mathcal{K}_n$ by imposing $s_{i,j} = - c_{i,j}$ for positive constants $c_{i,j}$ as well as
\begin{equation}
	X_{i,j} \geq 0 \quad \text{for all}\quad1\leq i < j \leq n\,.
\end{equation}
The such defined region $\mathcal{A}_{n-3}$ is a polytope -- more precisely an associahedron -- of dimension $(n-3)$. The $n$-point amplitude of biadjoint scalar theory is then given as the volume of the dual of this associahedron.

Associahedra are ubiquitous objects appearing in many contexts in mathematics. For example, the cluster polytopes of the cluster algebras of $\G{2,n}$ are equivalent to $\mathcal{A}_{n-3}$. Furthermore, the fan of the totally positive tropical configuration space $\TPTC{2,n}$ is equivalent to the dual of $\mathcal{A}_{n-3}$ \cite{Speyer2005}. We can thus compute the volume of the dual of $\mathcal{A}_{n-3}$ and in this way the $n$-point amplitude of biadjoint scalar theory by triangulating $\TPTC{2,n}$.

The biadjoint scalar amplitudes associated to $\TC{k,n}$ were generalized in \cite{Cachazo2019,Cachazo:2019apa} from $k=2$ to general values of $k$ by a generalization of the scattering equations from $\mathbb{C}\mathbb{P}^{1}$ to $\mathbb{C}\mathbb{P}^{k-1}$, for more recent work in this direction see also \cite{Sepulveda:2019vrz,Borges:2019csl,Cachazo:2019ble,Early:2019zyi}. We can hence compute these generalized biadjoint scalar amplitudes by analysing $\TC{k,n}$ or its totally positive part, which corresponds to the amplitude with canonical ordering.

As was demonstrated in \cite{Drummond:2019qjk}, using the fact that the cluster fan triangulates the fan of the totally positive tropical configuration space, we can compute its volume and obtain an expression for the generalised biadjoint scalar amplitude. Given a simplicial fan $\text{Tri}F_{k,n}$ triangulating the fan of the totally positive tropical configuration space $\TPTC{k,n}$, the amplitude is given by
\begin{equation}
	\label{equ:GBSA}
	m_k^n(\unit|\unit) = \sum_{\substack{\text{max. cones}\\[2pt]C\in \text{Tri}F_{k,n}}}\, \prod_{\substack{\text{rays}\\\vect{g} \in C}} \frac{1}{y\cdot\trop\Phi(\vect{g})}\,,
\end{equation}
whereas $\trop\Phi$ denotes the embedding of the totally positive tropical configuration space into the full tropical configuration space obtained via the web-parameterisation as given by eq.~\eqref{equ:webParam} and $y$ is the vector of lexicographically ordered generalized Mandelstam invariants $s_{i_1\dots i_k}$ spanning the kinematic space. In the case where $\text{Tri}F_{k,n}$ coincides with the cluster fan, as was previously considered, $C$ amounts to a cluster of the cluster algebra, and $\vect{g}$ to a ray associated to an $\A$-variable. With the top-dimensional cones being simplicial they each consist of $d$ rays, the dimension of the fan.

In this chapter we build on the work of \cite{Drummond:2019qjk} to analyse $\TPTC{3,8}$ and its properties. Using cluster algebras, we will compute a triangulation of its fan and obtain its full geometry, whose combinatorial structure is expressed in terms of its $f$-vector. Furthermore, we will show how to obtain a minimal, non-redundant triangulation which allows us to write down the generalised biadjoint scalar amplitude with the least amount of spurious poles. Finally, we briefly mention how this can be extended to obtain the amplitude associated to $\TPTC{4,8}$.

\subsection{Mutating and fusing cones}
In \cite{Herrmann2014} the question was posed, whether it is feasible to compute the tropical Grassmannian $\TG{3,8}$. Using cluster algebra techniques,  at least computing the totally positive tropical configuration space $\TPTC{3,8}$ becomes not only feasible but also efficient.

\begin{figure}[ht]
	\centering
	\includegraphics[width=0.5\textwidth]{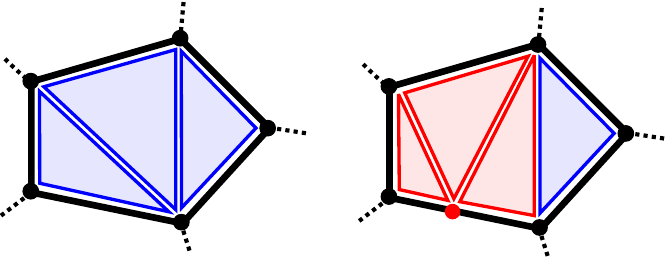}
	\caption{Illustrative example of a non-minimal triangulation of a non-simplicial cone. In this example, we consider a $3$-dimensional fan and draw its intersection with the unit-sphere $S^2$. Note that the vertices correspond to the rays and lines to dimension-$2$ surfaces. Simplicial cones have three rays and thus three vertices. On the left hand, we draw in blue a minimal triangulation. On the right hand, we draw how a redundant ray, drawn in red, leads to a non-minimal triangulation.}
	\label{fig:nonSimplicialRedundantTriangulation}
\end{figure}

Starting from the seed of the cluster algebra of $\G{3,8}$, we perform all possible mutations to obtain the full cluster fan. As stated in \cite{Drummond:2019qjk}, this contains $128$ rays, $8$ of which are redundant. These rays arise from redundant triangulations, as for example from the triangulation of an already simplicial cone by smaller simplicial cones. However, we also observe such redundant triangulations in (parts of) non-simplicial cones with up to $13$ rays. Such a scenario is illustrated in figure \ref{fig:nonSimplicialRedundantTriangulation}.

Proceeding as for $\G{4,8}$, we can also compute the truncated cluster algebra, that is, we restrict mutation to those clusters, that contain only non-redundant rays. Therefore, we truncate the cluster algebra whenever we encounter a cluster with at least one redundant ray. As demonstrated in \cite{Drummond:2019qjk}, this results in a truncated cluster fan with $120$ rays and $21720$ cones. This cluster fan, however, has holes resulting from the truncation. Thus, while it contains only non-redundant triangulations, it does not triangulate the full totally positive tropical configuration space $\TPTC{3,8}$.

As the cluster algebra is finite in this case, we can start from its full -- and finite -- fan and fuse the cones along the redundant faces to obtain $F_{3,8}$. In this way, we were able to construct the totally positive tropical configuration space efficiently using the associated cluster algebra. The $f$-vector of this and of the cluster fans are stated in table \ref{tab:gr38FVectors}.
\begin{table}[ht]
	\centering
	\begin{tabular}{|l|cccccccc|}
		\hline \hline
		Geometry & \multicolumn{8}{|c|}{$f$-vector} \\ \hline \hline
		Full cluster & $128$ & $2408$ & $17936$ & $67488$ & $140448$ & $163856$ & $100320$ & $25080$  \\ \hline
		Truncated cluster & $120$ & $2240$ & $16584$ & $61920$ & $127568$ & $146944$ & $88560$ & $21720$ \\ \hline
		\small $\TPTC{3,8}$& $120$ & $2072$ & $14088$ & $48544$ & $93104$ &  $100852$ & $57768$ & $13612$ \\ \hline \hline
	\end{tabular}
	\caption{$f$-vectors of the full and truncated cluster polytope as well as that of $\TPTC{3,8}$.}
	\label{tab:gr38FVectors}
\end{table}

\subsection{Minimal triangulation}
In the case of $\TPTC{3,8}$ the full cluster algebra is finite, resulting in a complete triangulation. However, the cluster fan has $8$ redundant rays resulting in parts of the totally positive tropical configuration space to be redundantly triangulated. 

By comparing the cluster fans of the full and truncated algebra, we identify $8$ holes in the truncated fan that correspond to the areas, where the cluster algebra produces redundant triangulations and was thus truncated. Fusing the cones of the full cluster algebra that lie in any of these holes, we can identify those cones of $F_{3,8}$ that are not yet minimally triangulated by the cluster algebra.\footnote{Note that in higher dimensions, even when excluding redundant rays, it is possible to obtain triangulations with different numbers of simplices. While all of them are non-redundant, only one of them is truly minimal.} 

Note that in this way we do not always find the complete non-minimally triangulated cone but only that part of the cone, that is not yet minimally triangulated. This can also be seen on the right hand side of figure \ref{fig:nonSimplicialRedundantTriangulation}, where the right part of the cone is non-redundantly triangulated by the cluster algebra and hence only the left part, drawn in red, actually lies inside the hole. In an abuse of language, we will nonetheless refer to this as the full cone.

\begin{figure}[ht]
	\centering
	\includegraphics[width=0.3\textwidth]{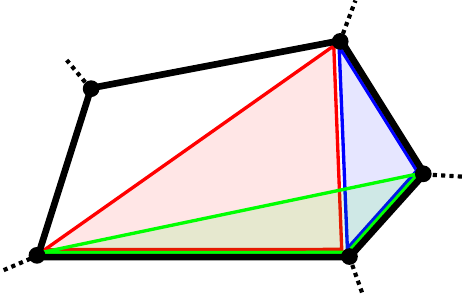}
	\caption{Illustrative example of the algorithm to triangulate a non-simplicial cone. Here we consider a $3$-dimensional fan and draw its intersection with the unit-sphere $S^2$. The non-simplicial cone of the totally positive tropical configuration space is drawn in black. The red, green and blue cones are part of the set of all cones that may be used for the triangulation. If the algorithm has already picked the red one for the triangulation, only the blue one can be picked in the next step, as the green one intersects with the red cone and does not share a codimension-$1$ boundary with it.}
	\label{fig:minimalTriangulation}
\end{figure}

Given any cone of $F_{3,8}$ we construct its minimal triangulation by first building all possible simplicial cones out of the rays of the non-simplicial cones (these are linearly independent combinations of $8$ rays, which is equal to the dimension of the fan, as may be seen by specialising eq.~\eqref{eq:DimGrkn} to our case). We then iteratively combine these cones to a triangulation by grouping them such that they share codimension-$1$ boundaries and do not intersect. This is illustrated in figure \ref{fig:minimalTriangulation}. When no such cones are left, we check whether all codimension-$1$ boundaries are shared between precisely two cones to verify whether the triangulation covers the entire cone. Finally, we simply choose the minimal triangulation out of the thus constructed ones.

Note that this algorithm is only applicable if there are no rays of the totally positive tropical configuration space inside of the hole of the truncated cluster algebra. For $\TPTC{3,8}$ this is indeed the case, as can be checked by comparing with the full cluster algebra, which is known to provide its complete triangulation. Due to the complexity of the problem, this algorithm is, however, not efficient for the largest non-simplicial cones encountered in these holes. In each of the $8$ holes we find one non-simplicial cone consisting of $13$ non-redundant rays which is redundantly triangulated by $24$ simplicial cones of the cluster algebra. For this cone the algorithm does not terminate within a reasonable amount of time.

In combination with the triangulation obtained from the truncated cluster algebra, we thus obtain a triangulation of the fan of $\TPTC{3,8}$ with $23496$ simplicial cones. More precisely, we find that, not considering the largest non-minimally triangulated cones, there are $3168$ simplicial cones corresponding to a redundant triangulation in the full cluster fan, $396$ for each of the $8$ holes. These cones redundantly triangulate $99$ non-simplicial cones (or parts of such) in each hole. In the minimal triangulation, these non-simplicial cones are triangulated by $198$ simplicial cones. Compared to the redundant triangulation, the amount of simplicial cones per non-simplicial cone is reduced by half.

Extrapolating this data to the largest non-minimally triangulated cones, which are redundantly triangulated by $24$ simplicial cones, would suggest that they have a minimal triangulation with $12$ simplicial cones. In this case, the minimal triangulation fan of $F_{3,8}$ would consist of $23400$ simplicial cones.

\subsection{Generalised biadjoint scalar amplitude}
In \cite{Drummond:2019qjk}, the simplicial fan of the full cluster algebra of $\G{3,8}$ was used to compute the respective generalised biadjoint scalar amplitude amplitude. However, having obtained the near-minimal triangulation, which consists of $1584$ cones less than the full cluster triangulation, the amplitude can be written in a more economic way. The full result is contained in the ancillary file \texttt{Gr38AmpMin.m} attached to the \texttt{arXiv} submission of this article.

We demonstrate how the amplitude can be written in a shorter form when passing to the minimal triangulation by considering the example of a simplicial cone that is redundantly triangulated. The cluster algebra triangulates this cone with two also simplicial cones containing one of the redundant rays. This redundant triangulation leads to the contribution 
\begin{align}
	&\quad 1/\Big(\left(y\cdot b_{3,12345678}\right)\left(y\cdot b_{3,78123456}\right)\left(y\cdot b_{4,12345678}\right)\left(y\cdot b_{5,34215678}\right)\left(y\cdot b_{5,67548123}\right) \nonumber\\
	&\qquad\quad \times \left(y\cdot b_{10,24513678}\right)\left(y\cdot b_{8,78564123}\right)\left(y\cdot b_e\right)\Big) \nonumber\\
	&+ 1/\Big(\left(y\cdot b_{3,12345678}\right)\left(y\cdot b_{3,78123456}\right)\left(y\cdot b_{4,12345678}\right)\left(y\cdot b_{5,34215678}\right)\left(y\cdot b_{5,67548123}\right) \nonumber\\
	&\qquad\quad \times \left(y\cdot b_{10,24513678}\right)\left(y\cdot b_{8,12345678}\right)\left(y\cdot b_e\right)\Big)\,,\label{eq:Gr38AmpRedundant}
\end{align}
where the $\R^D$ vectors $b$ correspond to certain rays of $F_{3,8}$. Their precise form may be found in \cite{Drummond:2019qjk}, however it will not be important for our purposes. The same contribution in our minimal triangulation reads
\begin{align}
	1/&\Big(\left(y\cdot b_{3,12345678}\right)\left(y\cdot b_{3,78123456}\right)\left(y\cdot b_{4,12345678}\right)\left(y\cdot b_{5,34215678}\right)\left(y\cdot b_{5,67548123}\right)\nonumber \\
	&\times \left(y\cdot b_{10,24513678}\right)\left(y\cdot b_{8,12345678}\right)\left(y\cdot b_{8,78564123}\right)\Big),\label{eq:Gr38AmpNonRedundant}
\end{align}
and indeed, \eqref{eq:Gr38AmpRedundant} simplifies to \eqref{eq:Gr38AmpNonRedundant} upon taking the common denominator, and using the following identity
\begin{equation}
	b_e = b_{8,12345678} + b_{8,78564123}\,,\label{eq:Gr38RayIdenity}
\end{equation}
that the redundant ray $b_e$ obeys. In general, the terms coming from the triangulation of non-simplicial cones are more complex and result, even for the minimal triangulations, in more than one term. These terms, however, cannot be further simplified while keeping the same structural form.

The advantage of writing the amplitude in this way is that such a minimal triangulation introduces the least possible amount of spurious poles into the amplitude. As can be seen from the simplification of the terms above, when using the minimal triangulation, the pole associated to $\left(y\cdot b_e\right)$, which in the redundant triangulation cancels between the two fractions, is not present at all. 

In general, we see from eq.~\eqref{equ:GBSA} that every ray introduces an apparent pole to the amplitude. The rays, which are dimension-$1$ faces of the triangulation fan, correspond to codimension-$1$ faces of the associated polytope in kinematic space. In this sense, if the cluster algebra introduces a redundant ray to triangulate the totally positive tropical configuration space, it introduces a codimension-$1$ boundary in the kinematic space that is associated to a non-physical pole whose contributions cancel between the terms.

\subsection{A note on $\TG{4,8}$}
For $\TPTC{3,8}$ we obtained a near-minimal triangulation by using the full $\G{3,8}$ cluster algebra, which in this case is finite. However $\G{3,n}$ and $\G{4,n}$ cluster algebras are no longer finite for $n\ge 9$ and $n\ge 8$, respectively. Therefore we cannot rely on them to obtain a triangulation of the entire totally positive tropical configuration space.

%Furthermore, as explained above, there are strong indications that even the fan of the entire infinite cluster algebra is not complete, that is there are areas that are not triangulated at all.

A potential alternative starting point is the truncated cluster algebra, in the sense discussed in section \ref{sec:tptg48}, where we also obtained the latter for $\G{4,8}$. Based on this result, in this subsection we will apply our general algorithm for eliminating redundant boundaries, and attempt to triangulate $\TPTC{4,8}$.

\begin{figure}[ht]
	\centering
	\includegraphics[width=0.6\textwidth]{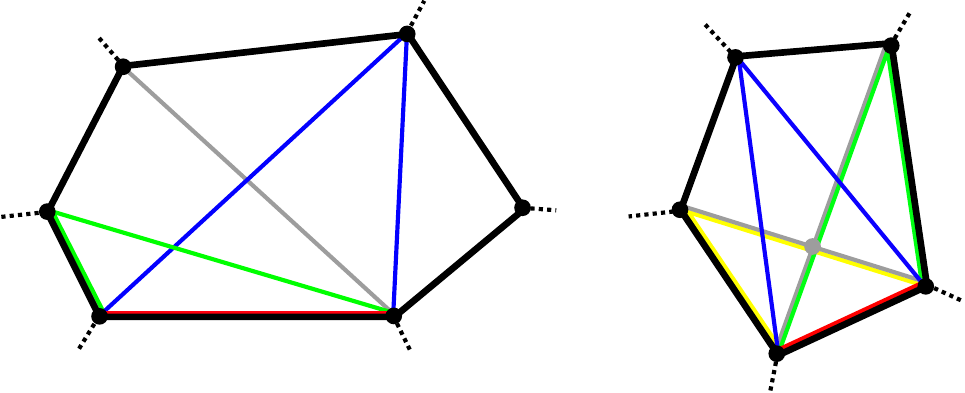}
	\caption{Illustrative examples of partially triangulated $3$-dimensional fans whose intersections with the unit-sphere $S^2$ are depicted. The hole in the truncated cluster fan is drawn in black. On the left hand side, the hole contains two non-simplicial cones separated by the grey line, representing a tropical hypersurface. The algorithm attempts to complete the faces on the boundary of the hole to simplicial cones. For the red face, two such possible completions are drawn in green and blue. Out of the two, only the green is valid, as the interior of the blue cone intersects the grey tropical hypersurface. On the right hand side we have two intersecting tropical hypersurfaces inside the hole, implying that it contains a valid ray. In this case, the red face cannot be completed to a valid cone by our algorithm.}
	\label{fig:triangulatingCone}
\end{figure}

Using the fact that any codimension-$1$ face is shared between precisely two clusters, we can first identify the codimension-$1$ faces on the boundary of each of the holes of the truncated cluster fan. Using any of the rays surrounding a given hole, we attempt to complete such a face to a simplicial cone, as is illustrated in figure \ref{fig:triangulatingCone}. For this, we first look at all possible such cones and then eliminate those that intersect any of the other cones or whose interior intersects a tropical hypersurface.\footnote{Note that using the tropical hypersurface equations obtained from the tropical parameterisation of the Plückers, eq.~\eqref{equ:webParam}, we may check whether any tropical hypersurface intersects the interior of a cone even though solving the equations for the hypersurfaces may not be efficiently possible.} Proceeding like this, we may obtain a valid, non-redundant though not necessarily minimal triangulation of the hole, which in turn may be used as a starting point to obtain the minimal triangulation.

%Having obtained the truncated cluster algebra of $\G{4,8}$ we can thus use this triangulation algorithm to also compute the amplitude associated to $\TPTC{4,8}$. 
In contrast to $\TPTC{3,8}$, for $\TPTC{4,8}$ we find only one hole in the truncated cluster fan whose boundary includes all $356$ rays. Furthermore, trying to complete the faces on the boundary of the hole to valid simplicial cones inside the hole only works for some of the faces causing the algorithm to stop without having triangulated the entire hole. This shows that the truncated cluster algebra has not produced all rays of $F_{4,8}$ and that the hole contains further non-redundant, valid rays. Such a scenario is illustrated on the right hand side of figure \ref{fig:triangulatingCone}.

We expect that a similar phenomenon will be present in all infinite cluster algebras, namely their truncations will not contain all rays of the corresponding tropical Grassmannians. Nevertheless, so far in our discussion we have not taken into account the rays that can obtained from limits of infinite mutation sequences, of the type we explored in section~\ref{sec:clusterSeq}. While these are not strictly part of the cluster algebra, starting from the latter they can be arrived at by a well-defined procedure, at least for the infinite mutation sequences associated with an affine $A_1^{(1)}$ subalgebra, as depicted in figure~\ref{fig:gr48AffineA2Cluster} for the case of $\G{4,8}$.

It remains to be seen whether (possibly more general) infinite mutation sequences can generate all $\TPTC{k,n}$ rays from any infinite $\G{k,n}$ cluster algebra. Within this class, we certainly anticipate $\G{4,8}$ and $\G{3,9}$ to be of similar complexity, and below the rest. This is because they have finite mutation equivalence class \cite{2006math......8367F}, namely the number of their different unlabeled quiver graphs or adjacency matrices is finite, even though an infinite number of cluster variables may be assigned to them. Whenever all rays of the tropical Grassmannian are generated by the infinite mutation sequences, then the proposal of~\cite{Drummond:2019qjk} for the computation of (canonically ordered) generalized biadjoint scalar amplitudes will still apply, and more generally the latter will always be equal to the volume of the positive tropical Grassmannian.

\section{Conclusion \& Outlook}\label{sec:Conclusions}
In this paper, extending the results of~\cite{Drummond:2019qjk}, we demonstrated how cluster algebras can be utilized as a very efficient tool to compute and understand the positive tropical configuration space $\TPTC{k,n}$. Making use of the simple algebraic mutation rules to generate cluster algebras, we obtained triangulation fans for $\TPTC{3,7}$, $\TPTC{3,8}$ and (parts of) $\TPTC{4,8}$ obtaining their entire geometries in accordance with previous results in the literature. On the other hand, this close connection between tropical geometry and cluster algebra allowed us to obtain results on the scattering amplitudes that are associated to either of these two objects.

In the case of $\TPTC{4,7}$, we used the cluster algebra to obtain a triangulation fan of the totally positive tropical configuration space as well as the $f$-vectors of these geometries in accordance with the results of \cite{Speyer2005}. Furthermore, we analyzed the cluster adjacency properties of both MHV and NMHV seven-particle amplitudes of $\N = 4$ SYM. Whereas for the MHV amplitudes we found that the totally positive tropical configuration space might explain some of the beyond--cluster adjacency properties, the number of pairs of consecutive letters in the symbol of the NMHV amplitudes saturates the $399$ dictated by cluster adjacency.

For $\TPTC{4,8}$, we used the totally positive tropical configuration space to obtain a finite subset of the infinite cluster algebra of $\G{4,8}$ by truncating the algebra at clusters containing at least one redundant ray corresponding to redundant triangulations. In agreement with \cite{Drummond:2019qjk}, this resulted in $356$ $\A$-variables arranged in $ $ clusters. Using this we made a proposal for the rational eight-particle alphabet of eight-point $\N=4$ SYM amplitudes.

Further to the rational letters appearing in the eight-particle symbol alphabet, we also discussed how the non-rational letters containing square roots found in \cite{Prlina2018,Zhang:2019vnm} may be obtained from limits of mutation sequences within infinite cluster algebras. We generalized the result of \cite{Canakci2018}, where the limit of certain mutation sequences in the affine rank-$2$ cluster algebra was discussed, to the affine rank-$2$ cluster algebra with principal coefficients. 

Finally, we showed how the cluster algebra can be used to obtain a minimal triangulation of $\TPTC{3,8}$ thus allowing us to write down the associated generalised biadjoint scalar amplitude with the near-minimal amount of spurious poles. Furthermore, making use of this triangulation, we computed the geometry of the totally positive tropical configuration space $\TPTC{3,8}$, expressed in terms of its $f$-vector.

This work can be extended in several directions. The rational letters that have been found previously in \cite{Prlina2018,Zhang:2019vnm} are contained in our proposal for the alphabet. In how far the additional rational letters that we have found, for example even some single Plücker variables like $\pl{1357}$ are not contained in their alphabet, play an actual physical role is still an open question. 

Furthermore it has to be seen whether limits of mutation sequences in the cluster algebra of $\G{4,8}$ reproduce the algebraic letters also found in \cite{Prlina2018,Zhang:2019vnm}. Further to their potential applications to the algebraic letters, it would also be very interesting to see whether and how similar results about mutation sequences can also be obtained in different or even general cluster algebras. Whereas first numerical tests indicate that these results can be extended to some affine rank-$3$ cluster algebras as well as the other affine rank-$2$ cluster algebra of $A_2^{(2)}$ Dynkin type, it seems that the ratio of $\A$-variables in the cluster algebra of hyperbolic type given by the adjacency matrix of eq.~\eqref{equ:rank2AdjM} with $b=c=3$ diverges. Finally, also the exact relation to the formalism put forward in \cite{Reading2018a,Reading2018b}, which leads to the same result, remains to be seen.

With respect to the geometry of $\TPTC{4,8}$ it would be interesting to obtain its triangulation, also to be able to write down the associated generalised biadjoint scalar amplitude. For this, however, the rays inside the hole that are not captured by the finite part of the cluster algebra have to be obtained. One approach could be to make use of the limits of mutation sequences.

Furthermore, the triangulation of polytopes has received some attention in mathematics starting at the end of the last century, see eg. \cite{Ziegler1995} and references therein. It would be interesting to see how the technology developed there can be used to obtain the triangulations computed in this paper in a more efficient way.

Finally, exploring whether  $\TPTC{4,9}$ may also be studied by similar means would be another exciting research direction for the future. It should be noted, however, that a great jump in complexity is expected compared to the cluster algebras of $\G{4,n}$ for $n=8$: The latter is of finite mutation type, which means that the topologies of the quivers that appear are restricted to a finite set, even if the number of clusters is infinite. This will no longer be the case for $n\ge 9$.

\section*{Acknowledgements}
We thank James Drummond, Jack Foster, {\"O}mer G{\"u}rdo{\u{g}}an and Marcus Spradlin for many stimulating discussions. NH is grateful for the support of the graduate programme of the Studienstiftung des deutschen Volkes.

\appendix

\section{Web-parameterisation of the totally positive configuration space}
\label{sec:webParam}
Following \cite{Speyer2005}, we construct a parameterisation of the totally positive part of the configuration space $\Conf{k,n}$. For this, we first draw the web graph $\text{Web}_{k,n}$ as follows. 

Starting with a $k$ by $n-k$ grid, with $k$ ingoing edges from the left and $n-k$ outgoing edges on the top, we label these external edges from $1$ to $n$ clockwise starting on the bottom left. We furthermore label the $d=(k-1)(n-k-1)$ internal chambers starting from the top left by filling the columns with $x_1,\dots,x_d$. The horizontal and vertical internal edges are all directed, pointing to the right and top, respectively.

\begin{example}
	In figure \ref{fig:webExamples}, the web graphs $\text{Web}_{2,5}$ and $\text{Web}_{3,7}$ used to obtain a parameterisation for $\Conf{2,5}$ and $\Conf{3,7}$, respectively, are depicted.
	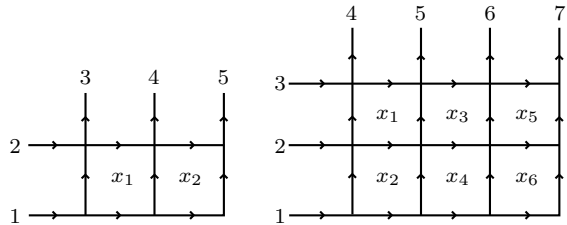
\begin{figure}[ht]
		\centering
		\begin{overpic}[scale=0.7,tics=5]{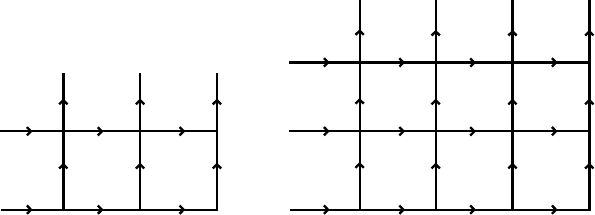}
			\put (15.5,7) {\scriptsize{$x_1$}}
			\put (28.1,7) {\scriptsize{$x_2$}}
			
			\put (-3.5,-0.6) {\scriptsize{$1$}}
			\put (-3.5,12.5) {\scriptsize{$2$}}
			\put (9.6,25.5) {\scriptsize{$3$}}
			\put (22.4,25.5) {\scriptsize{$4$}}
			\put (35.4,25.5) {\scriptsize{$5$}}
			
			\put (65,7) {\scriptsize{$x_2$}}
			\put (78,7) {\scriptsize{$x_4$}}
			\put (91,7) {\scriptsize{$x_6$}}
			\put (65,19) {\scriptsize{$x_1$}}
			\put (78,19) {\scriptsize{$x_3$}}
			\put (91,19) {\scriptsize{$x_5$}}
			
			\put (46,-0.6) {\scriptsize{$1$}}
			\put (46,12.5) {\scriptsize{$2$}}
			\put (46,24,4) {\scriptsize{$3$}}
			
			\put (59.3,37.5) {\scriptsize{$4$}}
			\put (72.2,37.5) {\scriptsize{$5$}}
			\put (85.2,37.5) {\scriptsize{$6$}}
			\put (98.2,37.5) {\scriptsize{$7$}}
			
		\end{overpic}
		\caption{Two examples for web graphs, $\text{Web}_{2,5}$ on the left and $\text{Web}_{3,7}$ on the right.}
		\label{fig:webExamples}
	\end{figure}
\end{example}

With the use of the web graph $\text{Web}_{k,n}$ we obtain a parameterisation of $\plAlt{i_1 \dots i_k}$ in terms of the web-variables $x_1,\dots,x_d$ by
\begin{equation}
	\plAlt{i_1 \dots i_k}\left(x_1,\dots,x_d\right) = \sum_{S\in\text{Path}\left(i_1,\dots,i_k\right)}\text{Prod}_S\left(x_1,\dots,x_d\right).
\end{equation}
The parts of this function are defined as follows. Given $K=\left(i_1,\dots,i_k\right)$ we construct all possible paths along the web graph in the following way. Denoting $[m]=\left(1,\dots,m\right)$, any such path may start at $[k]\setminus\left(K\cap[k]\right)$, that is at any of the ingoing edges except those whose label is contained in $K$. Furthermore, any path may end at $K\setminus\left(K\cap[k]\right)$, that is at any of the outgoing edges whose label is contained in $K$. The set $\text{Path}\left(K\right)$ is then given as all collections of non-intersecting paths such that all ingoing and outgoing labels that are allowed as start- and endpoints for the paths are covered by a path. Note that for $K=[k]$ there is no such path. However, in this case we include the empty set as a valid path.

\begin{example}
	An example for the collections of non-intersecting paths in $\text{Path}\left(2,5\right)$ is depicted in figure \ref{fig:pathsExample}.
	\begin{figure}[ht]
		\centering
		\begin{overpic}[scale=0.8,tics=5]{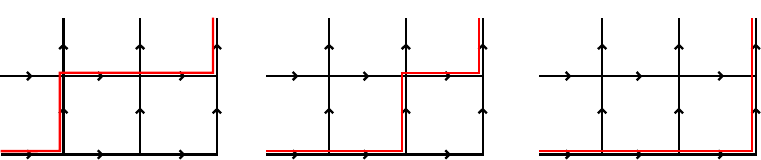}
			\put (12,5.3) {\scriptsize{$x_1$}}
			\put (22.5,5.3) {\scriptsize{$x_2$}}
			
			\put (-2.0,-0.1) {\scriptsize{$1$}}
			\put (-2.0,9.8) {\scriptsize{$2$}}
			\put (7.5,19) {\scriptsize{$3$}}
			\put (17.6,19) {\scriptsize{$4$}}
			\put (27.5,19) {\scriptsize{$5$}}
			
			\put (47,5.3) {\scriptsize{$x_1$}}
			\put (57.5,5.3) {\scriptsize{$x_2$}}
			
			\put (33,-0.1) {\scriptsize{$1$}}
			\put (33,9.8) {\scriptsize{$2$}}
			\put (42.5,19) {\scriptsize{$3$}}
			\put (52.6,19) {\scriptsize{$4$}}
			\put (62.5,19) {\scriptsize{$5$}}
			
			\put (83,5.3) {\scriptsize{$x_1$}}
			\put (93,5.3) {\scriptsize{$x_2$}}
			
			\put (69,-0.1) {\scriptsize{$1$}}
			\put (69,9.8) {\scriptsize{$2$}}
			\put (78.5,19) {\scriptsize{$3$}}
			\put (88.6,19) {\scriptsize{$4$}}
			\put (98.5,19) {\scriptsize{$5$}}
			
		\end{overpic}
		\caption{For $\text{Web}_{2,5}$ there are three complete collections of non-intersecting paths in $\text{Path}\left(2,5\right)$, each consisting of one path.}
		\label{fig:pathsExample}
	\end{figure}
\end{example}

For any complete collection of non-intersecting paths $S\in\text{Path}\left(K\right)$, we obtain the monomial $\text{Prod}_S\left(x\right)$ by multiplying for each path in $S$ the $x$ variables of those chambers, that are located above the path. This also includes those chambers that are not directly above the path but separated by other internal chambers. If there are no such chambers, we associate $1$ to the path.

\begin{example}
	We conclude our examples for $\text{Web}_{2,5}$ and $\text{Web}_{3,7}$. Consider the Plücker variable $\plAlt{25}$ of $\G{2,5}$. The three collections of non-intersecting paths are depicted in figure \ref{fig:pathsExample}. The path on the left hand side contributes by $x_1$, whereas the right hand path contributes by $1$. We thus obtain
	\begin{equation}
		\plAlt{25} = 1 + x_1 + x_1x_2\,.
	\end{equation}
	For the Plücker variable $\plAlt{356}$ of $\G{3,7}$, the collections of non-intersecting paths are depicted in figure \ref{fig:pathsExample2}. Summing over these collections, we obtain
	\begin{equation}
		\plAlt{356} = x_1^2x_2x_3x_4 + x_1x_2x_3x_4 + x_1x_2x_3\,.
	\end{equation}
	\begin{figure}[ht]
		\centering
			\begin{overpic}[scale=0.8,tics=1]{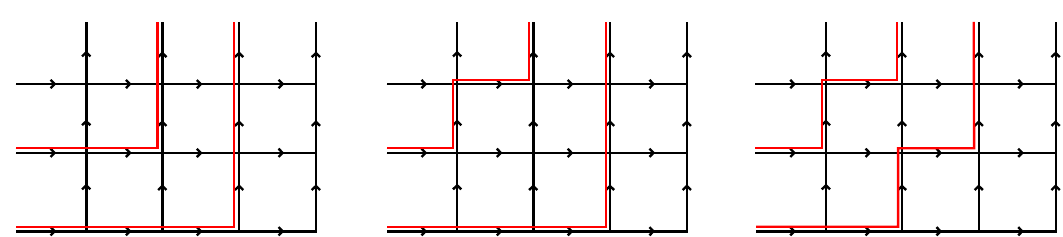}
				\put (10.8,4) {\scriptsize{$x_2$}}
				\put (17.9,4) {\scriptsize{$x_4$}}
				\put (25,4) {\scriptsize{$x_6$}}
				\put (10.8,10.5) {\scriptsize{$x_1$}}
				\put (17.9,10.5) {\scriptsize{$x_3$}}
				\put (25,10.5) {\scriptsize{$x_5$}}
				
				\put (-0.2,0) {\scriptsize{$1$}}
				\put (-0.2,7.2) {\scriptsize{$2$}}
				\put (-0.2,13.5) {\scriptsize{$3$}}
				
				\put (7.5,21) {\scriptsize{$4$}}
				\put (14.6,21) {\scriptsize{$5$}}
				\put (21.8,21) {\scriptsize{$6$}}
				\put (29.2,21) {\scriptsize{$7$}}
			
				\put (45.8,4) {\scriptsize{$x_2$}}
				\put (52.9,4) {\scriptsize{$x_4$}}
				\put (60,4) {\scriptsize{$x_6$}}
				\put (45.8,10.5) {\scriptsize{$x_1$}}
				\put (52.9,10.5) {\scriptsize{$x_3$}}
				\put (60,10.5) {\scriptsize{$x_5$}}
				
				\put (34.8,0) {\scriptsize{$1$}}
				\put (34.8,7.2) {\scriptsize{$2$}}
				\put (34.8,13.5) {\scriptsize{$3$}}
				
				\put (42.5,21) {\scriptsize{$4$}}
				\put (49.6,21) {\scriptsize{$5$}}
				\put (56.7,21) {\scriptsize{$6$}}
				\put (64.2,21) {\scriptsize{$7$}}
				
				\put (80.5,4) {\scriptsize{$x_2$}}
				\put (87.6,4) {\scriptsize{$x_4$}}
				\put (94.7,4) {\scriptsize{$x_6$}}
				\put (80.5,10.5) {\scriptsize{$x_1$}}
				\put (87.6,10.5) {\scriptsize{$x_3$}}
				\put (94.7,10.5) {\scriptsize{$x_5$}}
				
				\put (69.5,0) {\scriptsize{$1$}}
				\put (69.5,7.2) {\scriptsize{$2$}}
				\put (69.5,13.5) {\scriptsize{$3$}}
				
				\put (77.2,21) {\scriptsize{$4$}}
				\put (84.2,21) {\scriptsize{$5$}}
				\put (91.4,21) {\scriptsize{$6$}}
				\put (98.9,21) {\scriptsize{$7$}}
		\end{overpic}
		\caption{Collections of non-intersecting paths in $\text{Path}\left(3,5,6\right)$ of $\text{Web}_{3,7}$.}
		\label{fig:pathsExample2}
	\end{figure}
\end{example}

\section{Cluster algebras, coefficients and the $\g$-vector fan}
\label{sec:clusterAlgebras:coeffs}
The cluster fan as defined in section \ref{sec:clusterAlgebras:tropG} is closely related to the $\g$-vector fan defined for any cluster algebra in \cite{CAIV,GHKK2014}. In fact, as was stated in \cite{Drummond:2019qjk}, the mutation rule \eqref{equ:rayMutationRule} is a small modification of the mutation rule for the $\g$-vectors. To see the precise relation between these two fans, it is best to facilitate a different -- though equivalent -- perspective on cluster algebras and their mutation patterns. 

In this section, we will thus first review the construction of cluster algebras as laid out in \cite{CAIV} and finally propose a formula relating the two fans. Note that in difference to the main text of this paper, in this section we denote the rays associated to the $\A$-variables as $\g'$ and use $\g$ as originally introduced in \cite{CAIV}.

\subsection{Cluster algebras with coefficients}
As for example in the cluster algebras associated to $\G{k,n}$, we often deal with cluster algebras that have both, unfrozen and frozen nodes. Whereas the variables attached to the frozen nodes do not change under mutation, their dynamics is of interest in its own right and has many applications also in the context of theoretical physics \cite{CAIV,Fomin2003,Zamolodchikov1991}. Furthermore, analysing the dynamics of the frozen and unfrozen variables separately as well as their interplay allows a better understanding of both of them.

We thus promote the frozen variables to independent objects and introduce the coefficients $y_i$. Whereas the $\X$-variable associated to any unfrozen node is a monomial in both the frozen and unfrozen variables determined by the adjacency matrix, the coefficient attached to an $\A$-variable is a similar monomial but only in the frozen variables. 

We will use the following notation, adapting that of the mathematics literature to the more usual notation used in scattering amplitudes. The clusters of a cluster algebra of rank $r$ are labeled by an index $t$ such that a sequence of mutations as for example $\left(\mu_i,\mu_{i'}\right)$ results in a sequence of clusters denoted as $t \hsline{i} t' \hsline{i'} t''$. We denote the $\A$-variables of a cluster by $\vect{a}_t = \left(a_{1;t},\dots,a_{r;t}\right)$. The coefficients and $\X$-variables are similarly denoted by $\vect{y}_t$ and $\vect{x}_t$, respectively, whereas we will usually drop the label $t$ whenever it is clear from context to which cluster the objects belong.\footnote{Note that in the original mathematics literature, the $\A$-variables are usually denoted as $x_{i;t}$ whereas the $\X$-variables are labeled as $\hat{y}_{i;t}$.}

For the rank $d=(k-1)(n-k-1)$ cluster algebra of $\G{k,n}$, which is of \emph{geometric type}, we associate the triple $\left(\vect{a}_t,\vect{y}_t,B_t\right)$, the \emph{labeled seed}, to each cluster. The coefficients are defined for $1\leq j \leq d$ by
\begin{equation}
	y_{j;t} = \prod_{i=d+1}^{d+n} a_{i;t}^{b_{ij}^t}\,.
\end{equation}
in terms of the frozen variables $\left(a_{d+1;t},\dots,a_{d+n;t}\right)$. As usual, $B_t$ is the adjacency matrix with its components denoted by $b_{ij}^t$, which can be considered as a $(d+n)\times d$ matrix as there are no connections between frozen nodes. Note that these coefficients are closely related to the previously introduced $\X$-variables by
\begin{equation}
	\label{equ:coeffToX}
	x_{j;t} = \left(\prod_{i=1}^{d} a_{i;t}^{b_{ij}^t}\right)y_{j;t}\,.
\end{equation}

A special case of importance to the analysis of general cluster algebras is that of \emph{principal coefficients}. In essence, this means that for a rank $r$ cluster algebra we have $r$ frozen nodes, each connected to one of the unfrozen nodes, which can thus be identified with the coefficients $y_1,\dots,y_r$. This implies that the adjacency matrix of the initial seed is a $2r\times r$ matrix with the lower $r\times r$ part given by the identity matrix.

The advantage of associating both the variables and coefficients to each cluster and thus treating both on the same level is that the \emph{separation principle} becomes apparent. To see what this means, we first define the \textit{tropical addition}\footnote{Note that while this is closely related to the tropical addition discussed in section \ref{sec:tropicalGeometry} it is not quite the same!} $\oplus$ on the frozen variables for any real $b_i$ and $c_i$ by
\begin{equation}
	\label{equ:tropicalClusterAddition}
	\prod_{j=d+1}^{d+n} a_j^{b_j}\oplus\prod_{j=d+1}^{d+n}a_j^{c_j} = \prod_{j=d+1}^{d+n}a_j^{\min(b_j,c_j)}\,.
\end{equation}
With this addition, we can define the mutation for two labeled seeds $t\hsline{j}t'$ by
\begin{align}
	a_{j;t'} &= \frac{y_{j;t}\prod_{i=1}^da_{i;t}^{\left[b^t_{ij}\right]_+}+\prod_{i=1}^da_{i;t}^{\left[-b^t_{ij}\right]_+}}{\left(y_{j;t}\oplus 1\right)a_{j;t}}\label{equ:modClusterMutation}\,,\\\label{equ:modClusterMutation2}
	y_{l;t'} &= 
	\begin{cases}
		y_{j;t}^{-1}\quad&\text{if}\,l=j\\
		y_{l;t}y_{j;t}^{\left[b^t_{jl}\right]_+}\left(y_{j;t}\oplus 1\right)^{-b^t_{jl}}\quad&\text{if}\,l\neq j
	\end{cases}\,,
\end{align}
whereas the mutation for the adjacency matrix is as before and again $[x]_+=\max\left(0,x\right)$. In fact, when using the initial seed associated to the cluster algebra of $\G{k,n}$ this reproduces the previous results and is thus a completely equivalent notion of mutation.

\begin{example}
	We start with the initial seed associated to the quiver of figure \ref{quiv:gr25InitialQuiver}. That is, we have $\vect{a}_0 =\left(a_{1},a_2\right)$ and the initial coefficients and $\X$-variables are given by
	\begin{align}
		y_1 &= \frac{a_3a_6}{a_7}\,,\qquad x_1 = \frac{a_3 a_6}{a_2a_7} = \frac{1}{a_2}y_1 \label{eq:coeffEx1}\,,\\[3pt]
		y_2 &= \frac{a_5}{a_4a_6}\,,\qquad x_2 = \frac{a_1a_5}{a_4a_6} = a_1 y_2 \label{eq:coeffEx2}\,.
	\end{align}
	Performing all possible mutations, we obtain in total five seeds. Evaluating the addition $\oplus$ we can easily see that these are indeed the same results as by the usual mutation rule. For example, mutating the initial seed along $a_1$, we obtain a new variable given by
	\begin{equation}
		a'_1 = \frac{y_1+a_2}{a_1\left(y_1\oplus 1\right)}\,.
	\end{equation}
	To evaluate $y_1\oplus 1$ we specialize the addition of eq.~\eqref{equ:tropicalClusterAddition} to the case of eq.~\eqref{eq:coeffEx1} to obtain $y_1\oplus 1 = a_7^{-1}$ and get
	\begin{equation}
		a'_1 = \frac{a_3a_6+a_2a_7}{a_1}\,,
	\end{equation}
	which is exactly what one would get by using the original mutation rule \eqref{equ:clusterMutation}.
\end{example}

While this modified mutation rule is equivalent to the original formulation, it makes the separation of variables and coefficients directly apparent. To see this, first note that the monomial factor relating $x_{j;t}$ and $y_{j;t}$ can also be written as
\begin{equation}
	\prod_{i=1}^{d} a_{i;t}^{b_{ij}^t} = \left(\prod_{i=1}^{d} a_{i;t}^{\left[b_{ij}^t\right]_+}\right)\left(\prod_{i=1}^{d} a_{i;t}^{\left[-b_{ij}^t\right]_+}\right)^{-1}\,.
\end{equation}
Using this we can replace $y_{j;t}$ in the modified mutation rule \eqref{equ:modClusterMutation} by $x_{j;t}$ and thus obtain
\begin{equation}
	a_{j;t'} = \frac{x_{j;t}+1}{\left(y_{j;t}\oplus 1\right)}\cdot\frac{\prod_{i=1}^da_{i;t}^{\left[-b^t_{ij}\right]_+}}{a_{j;t}}\,.
\end{equation}

This property carries on even through multiple mutations such that any $\A$-variable can be written in such a separated form in terms of the $\A$-variables, coefficients and $\X$-variables of the initial seed. To be precise, given a cluster $\A$-variable $a$, there is a polynomial $F$ and integers $g_1,\dots,g_d$, such that we get
\begin{equation}
	\label{equ:gVect}
	a = \frac{F\left(x_1,\dots,x_d\right)}{F_\mathbb{T}\left(y_1,\dots,y_d\right)}a_1^{g_1}\dots a_d^{g_d}\,,
\end{equation}
where $F_\mathbb{T}$ denotes the same polynomial as $F$ with addition replaced by $\oplus$ and the variables and coefficients refer to the initial seed. In fact, the so-called $\g$-vector $\g=\left(g_1,\dots,g_d\right)$ is unique and therefore well-defined.

\begin{example}
	Turning again to our example of $\G{2,5}$, we can replace the coefficients in $a'_1$ by the $\X$-variables and thus obtain the factored form
	\begin{equation}
		\label{equ:aPrimeResult}
		a'_1 = \frac{x_1+1}{\left(y_1\oplus 1\right)}a_1^{-1}a_2\,.
	\end{equation}
	From this, we can immediately read off that $F(x_1,x_2) = 1+x_1$ and that the $\g$-vector of this variable is given by $\g=\left(-1,1\right)$.
\end{example}

Using the unique representation of the variables in eq.~\eqref{equ:gVect} we can define the $\g$-vector fan of the cluster algebra. To each cluster we associate the $\g$-vectors of its unfrozen variables and in this way obtain a cone. The $\g$-vector fan is then defined as the fan consisting of these cones \cite{Reading2018}. These fans are always simplicial and it is known that for infinite cluster algebras, they are not complete, meaning that the union of the cones of the fan does not cover the entire ambient space $\R^d$.

\subsection{Relation of $\g$-vector and cluster fan}
The thus defined $\g$-vectors are closely related to but not exactly the same as the rays of the cluster fan, here denoted by $\g'$. However starting from the unique representation of the $\A$-variables given in eq.~\eqref{equ:gVect} we can obtain another unique representation from which we obtain the rays of the cluster fan. Note that in this section, all $\A$- and $\X$-variables as well as the coefficients refer to those of the initial seed of the cluster algebra of $\G{k,n}$.

For this purpose we denote the tropical addition defined in eq.~\eqref{equ:tropicalClusterAddition} by $\T^-_a$, to exhibit that it is defined on the frozen variables with a minimum on the right hand side. Next, we extend it to the coefficients, $\X$-variables and their ratios $x_i/y_i$ for $1\leq i\leq d$ by the same formula except with a maximum instead of a minimum. We hence obtain for example the tropical addition $\T^+_{x/y}$ as
\begin{equation}
	\label{equ:modTropicalClusterAddition}
	\prod_{j=1}^d \left(\frac{x_j}{y_j}\right)^{b_j}\oplus\prod_{j=1}^d\left(\frac{x_j}{y_j}\right)^{c_j} = \prod_{j=1}^d\left(\frac{x_j}{y_j}\right)^{\max(b_j,c_j)}\,.
\end{equation}

To see how these modified additions are related to the original one, we consider an arbitrary polynomial $F$, as appearing in the unique representation of the $\A$-variables. We define the matrix $F_{ij}$ as the exponents appearing in the polynomial as
\begin{equation}
F\left(y_1,\dots,y_d\right) = \sum_m c_m\prod_{i=1}^d y_i^{F_{im}}.
\end{equation}
Using these definitions as well as the definition for the coefficients $y_j$ in terms of the frozen variables, we find that $F_{\T_y^+}$ and $F_{\T_a^-}$ are related by
\begin{equation}
	\label{equ:FtyToFTa}
	F_{\T^+_y}\left(y_1,\dots,y_d\right) = \left(\prod_{i=d+1}^{d+n} a_i^{\sum_{j=1}^d\left( b_{ij}\max_mF_{jm} - \min_m\left(b_{ij}F_{jm}\right)\right)} \right)F_{\T^-_a}\left(y_1,\dots,y_d\right)\,.
\end{equation}
This especially implies that the two expressions are related by a monomial in the frozen variables, which we denote by $C_F$. Furthermore, we can similarly see that the polynomials $F_{\mathbb{T}^+_x}$ and $F_{\mathbb{T}^+_y}$ are related by
\begin{equation}
	\label{equ:FtxtoFty}
	F_{\T^+_x}\left(x_1,\dots,x_d\right) = F_{\T^+_{x/y}}\left(\frac{x_1}{y_1},\dots,\frac{x_d}{y_d}\right)\cdot F_{\T^+_y}\left(y_1,\dots,y_d\right)\,.
\end{equation}
The ratios $x_i/y_i$ are given by monomials in the initial unfrozen variables, as can be seen from eq.~\eqref{equ:coeffToX}. In this way, we can directly evaluate the function $F_{\mathbb{T}^+_{x/y}}$ to be
\begin{align}
	&F_{\T^+_{x/y}}\left(\frac{x_1}{y_1},\dots,\frac{x_d}{y_d}\right) = {\sum_m}^\oplus \prod_{j=1}^d\left(\frac{x_j}{y_j}\right)^{F_{jm}} = \prod_{j=1}^d\left(\frac{x_j}{y_j}\right)^{\max_m(F_{jm})} \nonumber\\
	&\quad = \prod_{j=1}^d\left(\prod_{i=1}^{d} a_{i;t}^{b_{ij}^t}\right)^{\max_m(F_{jm})} = \prod_{i=1}^{d} a_{i;t}^{\sum_{j=1}^d b_{ij}\max_m(F_{jm})}\label{equ:Ftxy}\,.
\end{align}

Starting from the unique representation of the $\A$-variables given in eq.~\eqref{equ:gVect} and using the different tropical additions discussed in eqs.~\eqref{equ:FtyToFTa}, \eqref{equ:FtxtoFty} and \eqref{equ:Ftxy},\footnote{Note that the actual tropical addition \eqref{equ:tropicalClusterAddition} is defined on a certain semifield. In our modified definitions, this is not discussed at all. However, we can simply see these as formal manipulations that considered just by themselves do hold.} we therefore obtain the following unique representation
\begin{equation}
	\label{equ:aLetterRayForm}
	a = C_F^{-1} \frac{F\left(x_1,\dots,x_d\right)}{F_{\mathbb{T}^+_x}\left(x_1,\dots,x_d\right)}a_1^{g'_1}\dots a_d^{g'_d}\,,
\end{equation}
where the integer vector $\g'$ is related to the usual $\g$-vector by
\begin{equation}
	\label{equ:GtoCFan}
	g'_i = g_i + \sum_{j=1}^d b_{ij}\max_m F_{jm}\,.
\end{equation}

The remarkable property of this result, obtained by somewhat formal computations, is that, as can be and was checked for the finite and infinite cluster algebras discussed in this paper, the thus obtained vector $\g'$ associated to each $\A$-variable precisely coincides with the rays of the cluster fan as defined via the mutation rule \eqref{equ:rayMutationRule} in section \ref{sec:clusterAlgebras:tropG}.

\begin{example}
	Let us see how this plays out in our example from before. Due to simplicity, we will not directly use the formulas discussed before but demonstrate their general derivation on this example by performing the manipulations step by step. 
	
	We begin with the separated form of $a'_1$ given in eq.~\eqref{equ:aPrimeResult} and evaluate the tropical addition $(y_1\oplus 1)$. Using that $y_1 = a_3a_6a_7^{-1}$ we see that this evaluates to $F_{\T_a^-}\left(y_1,y_2\right)=a_7^{-1}$. In contrast, evaluating this for the modified addition we obtain $F_{\T_y^+}\left(y_1,y_2\right)=y_1 = a_3 a_6F_{\T_a^-}\left(y_1,y_2\right)$. Furthermore, we immediately see that $F_{\T_x^+}\left(x_1,x_2\right) = x_1 = a_2^{-1} F_{\T_y^+}\left(y_1,y_2\right)$ which ultimately leads to
	\begin{equation}
		a'_1 = \left(a_3a_6\right)^{-1}\frac{F\left(x_1,x_2\right)}{F_{\T_x^+}\left(y_1,y_2\right)}a_1^{-1}\,.
	\end{equation}
	
	We thus obtain the $\g'$-vector for this variable as $\g'_{a'_1} = \left(-1,0\right)$. It can easily be seen that this is exactly what we obtain by using the mutation rule for the rays given in eq.~\eqref{equ:rayMutationRule}.
\end{example}

\bibliographystyle{SupportingFiles/JHEP}
\bibliography{SupportingFiles/references}

\end{document}